\documentclass[journal]{IEEEtran}
\usepackage{amsmath,amsfonts}
\usepackage{algorithm}
\usepackage{array}
\usepackage[caption=false,font=normalsize,labelfont=sf,textfont=sf]{subfig}
\usepackage{textcomp}
\usepackage{stfloats}
\usepackage{url}
\usepackage{verbatim}
\usepackage{graphicx}
\usepackage{algpseudocode}
\usepackage{xcolor}
\usepackage{multirow}
\usepackage{cases}
\usepackage{amsthm}
\usepackage{amsmath}
\usepackage[normalem]{ulem}

\usepackage{accents}
\newcommand{\ubar}[1]{\underaccent{\bar}{#1}}

\theoremstyle{plain}
\newtheorem{theorem}{Theorem}

\newtheorem{assumption}[theorem]{Assumption}

\usepackage[english]{babel}
\usepackage[backend=biber,style=ieee,sorting=none]{biblatex}
\AtEveryBibitem{%
  \clearfield{issn} 
  \clearfield{issue} 
  \clearfield{isbn} 
  \clearfield{doi} 
    \clearlist{language} 
    \clearfield{urlyear} 
    \clearfield{note}   

  \ifentrytype{online}{}{
    \clearfield{url}
  }
}

\definecolor{mario}{rgb}{0.8,0.8,1}

\begin{document}

\title{Optimization-based Heuristic for Vehicle Dynamic Coordination in Mixed Traffic Intersections}

\author{Muhammad Faris, Mario Zanon\textsuperscript{*}, and Paolo Falcone
\thanks{\textsuperscript{*}Corresponding author. This work was partially supported by the Wallenberg Artificial Intelligence, Autonomous Systems, and Software Program (WASP) funded by the Knut and Alice Wallenberg Foundation.}
\thanks{M. Faris is with the Department of Electrical Engineering, Chalmers University of Technology, Gothenburg, Sweden (e-mail: farism@chalmers.se)}
\thanks{M. Zanon is with the IMT School for Advanced Studies Lucca, Lucca, Italy (e-mail: mario.zanon@imtlucca.it)}
\thanks{P. Falcone is with the Department of Electrical Engineering, Chalmers University of Technology, Gothenburg, Sweden, and Dipartimento di Ingegneria “Enzo Ferrari”, Universita di Modena e Reggio Emilia, Modena, Italy (e-mail: paolo.falcone@chalmers.se)}
}

\markboth{Journal of \LaTeX\ Class Files,~Vol.~14, No.~8, August~2021}%
{Shell \MakeLowercase{\textit{et al.}}: A Sample Article Using IEEEtran.cls for IEEE Journals}


\maketitle

\begin{abstract}
In this paper, we address a coordination problem for connected and autonomous vehicles (CAVs) in mixed traffic settings with human-driven vehicles (HDVs). The main objective is to have a safe and optimal crossing order for vehicles approaching unsignalized intersections. This problem results in a mixed-integer quadratic programming (MIQP) formulation which is unsuitable for real-time applications. Therefore, we propose a computationally tractable optimization-based heuristic that monitors platoons of CAVs and HDVs to evaluate whether alternative crossing orders can perform better.  It first checks the future constraint violation that consistently occurs between pairs of platoons to determine a potential swap. Next, the costs of quadratic programming (QP) formulations associated with the current and alternative orders are compared in a depth-first branching fashion. In simulations, we show that the heuristic can be a hundred times faster than the original and simplified MIQPs and yields solutions that are close to optimal and have better order consistency.           
\end{abstract}


\section{Introduction}
\label{sec:introduction}
%
\IEEEPARstart{T}{raffic} intersections, along with other merging areas such as on ramps or roundabouts, are widely recognized as significant bottlenecks in the road network, contributing to traffic-related issues that encompass both safety and inefficient use of the infrastructure. Intersection areas are prone to a considerable number of traffic accidents and fatalities \cite{choi_crash_2010}. To address safety concerns, strict traffic control measures are implemented, such as traffic lights and signs. Additionally, merging areas are characterized by frequent stop-and-go patterns, and the traffic flow is typically reduced by human drivers' behavior. These inefficiencies further lead to increased pollution and energy consumption \cite{silva_intersections_2022}. Achieving traffic efficiency while ensuring safety at all times poses a significant challenge that requires innovative approaches.

Traffic lights have long been a popular conventional approach to coordinating human-driven vehicles (HDVs), using a top-down perspective \cite{Sharon2017}. However, despite their widespread use, traffic lights alone often fail to fully resolve traffic problems and achieve the desired balance between traffic efficiency and safety. One of the issues with traffic lights is their low throughput, which leads to a build-up of vehicles waiting for their turn to occupy the intersection, resulting in traffic jams. Another issue is that human errors highly contribute to accidents in signalized intersections~\cite{AlejandroIvanMoralesMedina2020}.

To address these challenges and improve traffic management, connected and autonomous vehicles (CAVs) can play a vital role \cite{Hult2016}. They can leverage advanced communication and sensing technologies to unlock new possibilities for a more efficient traffic flow. By sharing real-time information and coordinating their actions through vehicle-to-everything (V2X) communication schemes, CAVs can ensure smoother traffic operations and better adherence to traffic rules than HDVs. 

As research on autonomous vehicles continues to grow, the primary focus is on gradually replacing human drivers' roles with automation and enhancing overall traffic performance. In the future, the development of CAVs is expected to enable the implementation of unsignalized intersections, where coordinated interactions between vehicles will replace traditional traffic signals~\cite{Mertens2022a}. 

The main challenge stemming from unsignalized intersections is priority assignment, i.e., selecting a crossing order for the incoming vehicles. This is a combinatorial problem, i.e., mixed-integer problem (MIP), which is NP-hard~\cite{chouhan_autonomous_2018} and whose computational complexity grows exponentially with the number of vehicles. This makes a real-time application of such MIP impossible. As an alternative, first-come first-serve (FCFS) scheduling has been widely used due to its simplicity. Unfortunately, this approach can be far from optimal \cite{meng_analysis_2018}. 

Before achieving full market penetration of CAVs, transition phases will take place, where HDVs and CAVs will share the road and interact with each other. Despite a possible future high penetration of CAVs, the presence of legacy HDVs cannot be entirely ruled out. As a result, a specialized coordination strategy for CAVs is essential to effectively handle interactions with HDVs, particularly concerning occupancy at unsignalized merging areas or intersections. This dedicated strategy must ensure safe and efficient traffic flow in such mixed traffic scenarios.

Accounting for the behavior of HDVs in coordination can be very complex, as decisions such as determining the crossing order and acceleration profile need continuous adaptation to account for the uncertain HDV trajectories. The presence of HDVs may trigger changes in order as the current order may become inefficient or even infeasible. Continuously monitoring and assessing the behavior of HDVs to close the control loop can be computationally challenging if the conventional approach of resolving the MIP problem is applied.

To address these issues, this work proposes a heuristic algorithm for vehicle crossing order problems in mixed traffic. The main focus is on handling the dynamic reordering of vehicles caused by changes in HDVs' predicted trajectories. The problem is first modeled and formulated as a mixed-integer quadratic programming (MIQP), where HDVs are grouped into mixed-platoons. The proposed heuristic is then derived as an approximation of the MIQP, inspired by branch-and-bound (B\&B) strategies but specifically tailored to this problem. 
Rather than performing a standard B\&B search, we exploit the knowledge of the problem to focus the expansion of the decision tree on a very small subset of all possible orders. This subset is selected based on potential future safety constraint violations, i.e., by a consistency check. At each time, an alternative crossing order that swaps two adjacent platoons is compared to the current one. The new order is retained if its cost is lower than that of the current order. 

In summary, this work makes the following primary contributions:

\begin{itemize}
    \item We formulate an MIQP optimal coordination problem based on the platooning strategy
    \item We propose an optimization-based heuristic to solve the problem in computationally tractable way
    \item We perform comprehensive numerical simulations that demonstrate the effectiveness of our approach
\end{itemize}

This paper is organized as follows. In Section~\ref{sec:introduction} we provide an extensive review of the literature related to the considered problem. In Section~\ref{sec:problem_statement} we define the problem more in detail and set the premises for the formulation of the optimal coordination, which is provided in Section~\ref{sec:coordination_problem}. In Section~\ref{sec:heuristic}, we introduce the heuristic algorithm. All simulations and discussions on the performance of the different algorithms are given in Section~\ref{sec:simulation}. Finally, we conclude the work in Section~\ref{sec:conclusions}. 

%
\section{Literature Review}
\label{sec:literature_review}
%
Early research on vehicle coordination at unsignalized intersections primarily focused on autonomous vehicles (AVs). In \cite{Makarem2013}, a coordination problem for AVs was proposed, suggesting the use of a safety distance between conflicting vehicles for collision avoidance. The occupancy priority was determined based on arrival times at the intersection. A similar problem was addressed by \cite{Campos2013}. Both acknowledged the computational complexity of the crossing order problem and proposed a reachability-based heuristic method as an alternative to mixed-integer programming (MIP) formulation.

In subsequent research, several alternative heuristic methods have been utilized. The reachability-based heuristic has been further developed into a sequential priority decision-making method in \cite{DeCampos2017}, which aimed at reducing the number of priority permutations. In~\cite{Hult2019a}, a heuristic based on mixed-integer quadratic programming (MIQP) was introduced as an approximation of vehicle coordination using timeslots. Other trajectory optimization-based methods using constraint programming can be found in \cite{Murgovski2015, Karlsson2018}, and \cite{deng_cooperative_2023}. To handle the complexity of NP-hard formulations, tree search methods were applied in \cite{xu_cooperative_2020} and \cite{Xu2020a} to explore possible crossing orders, finding optimal solutions through multiple directed iterations, even in lane change applications. Additionally, rule-based heuristics combined with timeslot or exit time minimization were developed in \cite{Mahbub2020} and \cite{chalaki_optimal_2019}. However, these heuristics are not specifically designed to address mixed traffic schemes. 

More recently, researchers have been extending their focus to mixed traffic scenarios, in particular involving human-driven vehicles (HDVs). Rule-based protocols were used alongside traffic lights by \cite{Sharon2017} and \cite{Aoki2019}. In \cite{Chen2020} the same problem was addressed using the CAV-HDV platoon control method. Platooning strategies were also explored in \cite{Peng2021} for intersection cases. A cooperative maneuver technique was proposed by \cite{Mertens2022a} for HDVs that are connected and behave according to a specific model. The crossing order was obtained through gradient-based optimization, although safety was not a primary concern for HDVs. A similar setting was considered in \cite{Shen2019}, albeit with First-Come-First-Serve (FCFS) priority. Furthermore, \cite{faris_optimization-based_2022} studied the impact of involving HDVs on performance and safety, emphasizing that uncertainty and prediction mismatches from the HDVs may necessitate a change of order.

Some studies focus specifically on dynamic reordering or reprioritization problems. For example, \cite{scheffe_increasing_2022} implemented a time-varying priority assignment by evaluating possible collisions from each vehicle. In \cite{molinari_real_time_2020}, a negotiation-based priority approach was applied to coordinate CAVs, allowing rules to be negotiated during the auction phase based on the current vehicle states. Arrival/exit time minimization-based methods were used in \cite{chalaki_priority_2022}, \cite{Xiao2020a}, and \cite{bifulco_decentralized_2022} to handle changing traffic flow. These methods sort the order based on individual assessments relying on conservative assumptions, such as current states only or maximum accelerations. However, they did not provide comparisons of solution quality and did not specifically address challenges in mixed traffic environments.

In this paper, we specifically deal with the challenges of dynamic reordering in the context of mixed traffic. The work begins by formulating MIQPs of the vehicle coordination, in original and simplified forms, used as a solution benchmark. Then, we propose an optimization-based (QPs) heuristic to address the computational challenge. We extensively evaluate the methods' performance in mixed traffic reordering scenarios, e.g., including when and which order to change. In addition, comparisons with the MIQPs are provided to assess the quality of the solutions and computational tractability. 

%
\begin{figure}[]
    \begin{center}
    \graphicspath{{./figures/}}
    \includegraphics[scale = 0.5]{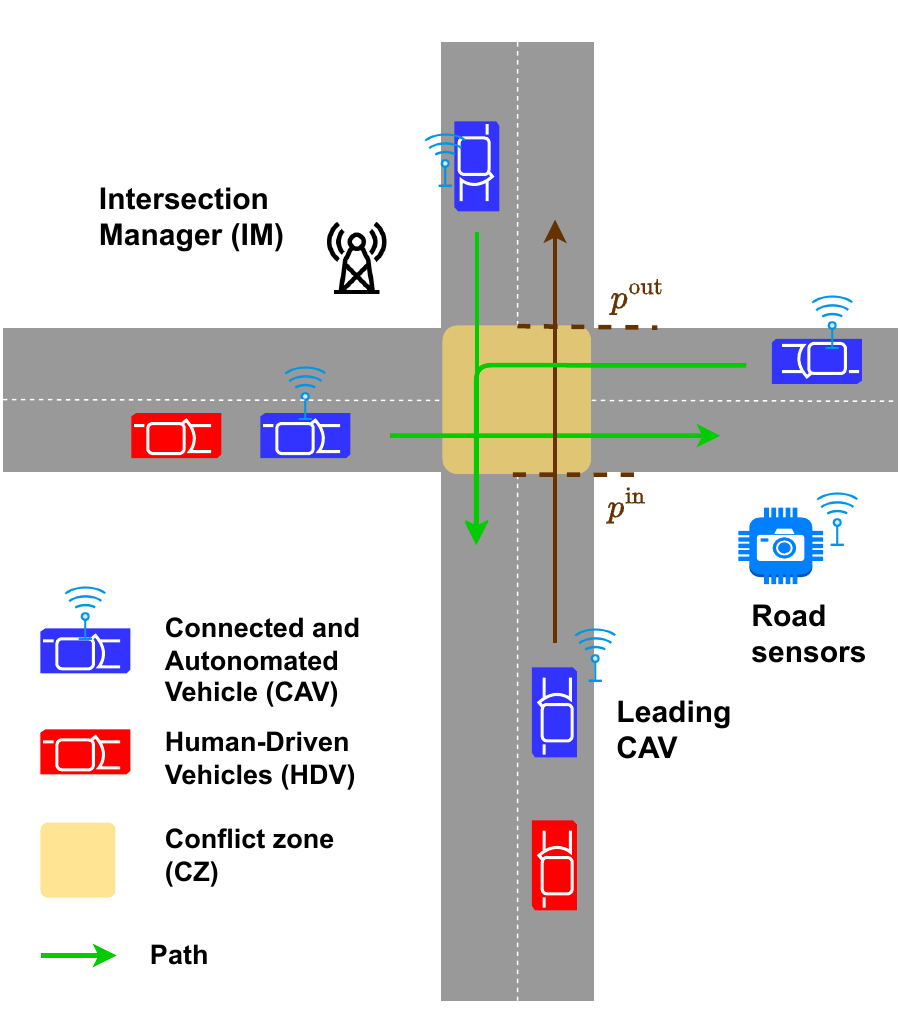}    
        \caption{Mixed traffic of CAVs and HDVs}
        \label{fig:mixed-intersection}
    \end{center}
\end{figure}
%

%
\section{Problem Setup \& Modeling}
\label{sec:problem_statement}

In this section, we introduce the problem setting and the notation used to describe both the intersection and vehicles.

\subsection{Types of Vehicles}

Let us consider a set of~$N+M$ vehicles, where~$N$ and~$M$ are, respectively, the numbers of \emph{connected and autonomous vehicles} (CAVs) and \emph{human-driven vehicles} (HDVs). Each vehicle is assigned an integer index, and we denote by $\mathcal{N},~\mathcal{M}$ the sets of indices relative to the CAVs and HDVs, respectively.

%
\subsection{Vehicle Modeling}

While any vehicle model can be used within our approach, in this paper we assume for simplicity that vehicle $i$ moves along its predefined path as described by a discrete-time, double-integrator model
\begin{equation}
    x_{i,k+1} = Ax_{i,k} + Bu_{i,k},  \qquad \forall i \in \mathcal{N}, \mathcal{M}, 
    \label{eq:dynamic}
\end{equation}%
where $k = \lfloor t/\Delta t \rfloor \in \mathbb{N}$ is the discrete-time index, $\Delta t$ is the periodic sampling time, $t \in \mathbb{R}_{+}$ is continuous-time instance, and
\begin{equation*}
    A = \begin{bmatrix}
    1 &  \Delta t \\
    0 & 1 
    \end{bmatrix}, \quad
    B = \begin{bmatrix}
    \frac{1}{2}\Delta t^{2} \\ \Delta t 
    \end{bmatrix}.
\end{equation*}
The state vector $x_{i,k} = \left[p_{i,k}, ~ v_{i,k}\right]^{\top}$ contains the longitudinal distance of vehicle $i$ from its origin point and its velocity. Each vehicle starts at $k=0$ with given initial states
\begin{align}
    x_{i,0} &= x^{\mathrm{0}}_{i}, \qquad \forall i \in \mathcal{N}, \mathcal{M}
\end{align}
%
%
and is subject to the following velocity and acceleration/deceleration (input) bounds
\begin{subequations}
    \begin{align}
        v^{\mathrm{min}} &\leq v_{i,k} \leq v^{\mathrm{max}}, \quad \forall i \in \mathcal{N},\mathcal{M},
        \label{eq:limv} \\
        u^{\mathrm{min}} &\leq u_{i,k} \leq u^{\mathrm{max}}, \quad \forall i \in \mathcal{N},\mathcal{M},
        \label{eq:limu}    
    \end{align}
    \label{eq:limvu}    
\end{subequations}
with $v^{\mathrm{min}}>0$, as we assume that vehicles can not reverse. 


The HDVs' behavior is described by the following mixed of constant-velocity and maximum acceleration/deceleration model
\begin{align}
    v_{i,k+1} &= v_{i,k} + \Delta t u_{i,k},~ \ k>0, ~\forall \ i \in \mathcal{M}, 
    \label{eq:hdv-cd}
\end{align}
where
\begin{align*}
    u_{i,k} = &\begin{cases}
    0, \quad &u_{i,k-1} \geq 0,  \\      
    u^{\mathrm{min}}, \quad &u_{i,k-1} < 0,
    \end{cases} 
\end{align*}
in which $u^{\mathrm{min}}$ is the deceleration limit.

\subsection{Conflict Zones}
We call a \emph{Conflict Zone} (CZ) any portion of the road intersection where vehicles coming from different directions intersect their paths, as illustrated in Figure~\ref{fig:mixed-intersection}. 

For simplicity, we consider the case of a single CZ at the center of the intersection area. The CZ is defined by the pairs $\left(p^{\mathrm{in}},p^{\mathrm{out}}\right)$ denoting the entry and exit positions in each direction, along each path. Furthermore, for simplicity, we consider the case of one lane per direction, which implies that no overtaking is allowed among vehicles coming from the same direction. 

To avoid lateral (side) collisions due to conflicting paths, vehicles must occupy the CZ exclusively, e.g. in an \emph{ordered} way, as it will be discussed in more details in Section~\ref{sec:coordination_problem}.

\subsection{Platooning roles}

In this paper, we aim at exploiting the presence of CAVs to (e.g., energy-) efficiently regulate the traffic at the intersection area. The main idea is that, by adapting the speed of CAVs approaching the intersection, the behavior of HDVs can be influenced to optimize the overall intersection efficiency and safety.

As stated in Assumption 1, an autonomous Intersection Manager (IM) is present, which assigns a CAV $i$ the role of \emph{platoon leader} if it approaches the intersection ahead of at least one HDV. We denote the last vehicle in each platoon $i$ as \emph{tail} $m$. The platoon length $l^\text{p}_{i,k}$ is defined as the position difference between the positions of the platoon leader and tail vehicles, respectively, i.e.,
\begin{align}
    l^\text{p}_{i,k} &= p_{i,k} - p_{m,k}.
\end{align}
Hence, a CAV that is not followed by any HDV forms a platoon with length $l^\text{p}_{i,k} = 0$ and is called \emph{one-vehicle platoon}.  

As HDVs do not have connectivity, they are not assigned any active role. 

\begin{assumption}
    HDVs cannot communicate with other vehicles or the IM. Nevertheless, their current states and inputs can be measured by the road infrastructure and are available to the IM. Moreover, a platoon must remain intact when crossing the intersection, i.e., no vehicle coming from other directions can divide the platoon.  
    \label{ass:cav-role}
\end{assumption}

Finally, we assume that a high penetration of CAVs is achieved, which we state as ${N} >> {M}$. This reduces the probability of having two or more HDVs coming to the intersection that are not preceded by any CAV. Note that this assumption is not necessary for the theory, but just to guarantee that our approach can yield some performance increase.

\section{Coordination Problem}
\label{sec:coordination_problem}

\subsection{Intersection Crossing Order}

Let $\mathcal{O}_t = [o_{1,k}, ..., o_{n,k} , ..., o_{N,k}]^{\top}$ be a vector containing the CAVs' crossing orders at time $t$, where $o_{n,k} = i \in \mathcal{N}$ if at time~$t$ CAV $i$ is planned to cross at the~$\iota$-th position. Accordingly, let $\iota_k(i)$ define a function yielding the crossing order~$\iota$ of CAV~$i$ in $\mathcal{O}_t$. Note that only the leading CAVs are ordered in $\mathcal{O}_t$, while the HDVs inherit the order assigned to the CAV leading their platoon.

\subsection{Safety Constraints}

In the considered problem setup, the vehicles approaching the intersection must avoid two types of collisions: lateral (side) collisions, which can occur in the CZ; and rear-end collisions, which can occur inside a platoon or between platoons that move in the same direction.

\subsubsection{Lateral collision avoidance}
\begin{figure}[]
    \begin{center}
    \graphicspath{{./figures/}}
    \includegraphics[scale = 0.6]{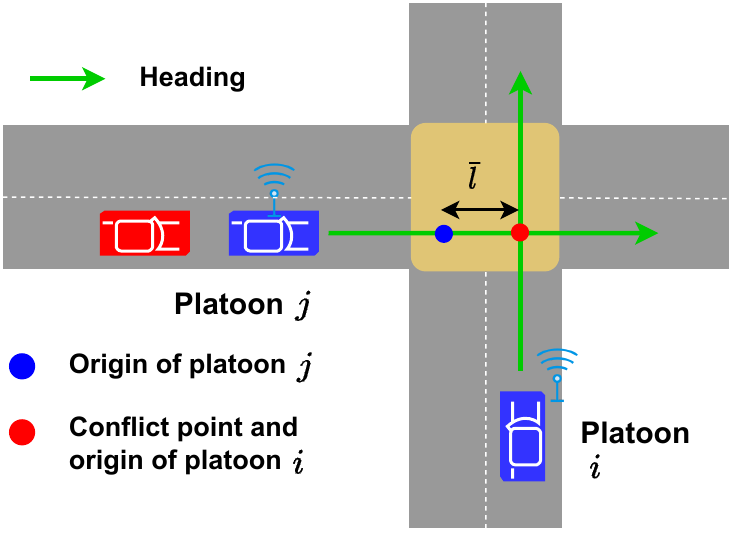}    
        \caption{Change of coordinate}
        \label{fig:ori-adjust}
    \end{center}
\end{figure}
To avoid side collisions within the CZ, the position gap between two platoons $i,j \in \mathcal{N}$ that arrive from different directions must be greater than a given constant $d^{\mathrm{min}}$. This collision avoidance condition is formulated within the following conditional constraint, which also defines the crossing order
\begin{subequations}
    \begin{align}
        \mathrm{if} ~\iota_k(j) &< \iota_k(i): \nonumber \\     
        p_{j,k} &\geq p_{i,k} + d^{\mathrm{min}} + \Bar{l}, ~ k \in [\ubar{k}_j,\bar{k}_i],      \label{eq:mip-sdca-a}          \\
        \mathrm{if} ~\iota_k(i) &< \iota_k(j): \nonumber \\        
        p_{i,k} &\geq p_{j,k} + d^{\mathrm{min}} + \Bar{l}, ~ k \in [\ubar{k}_i,\bar{k}_j], \label{eq:mip-sdca-b}       
    \end{align}%
    {\label{eq:mip-sdca}}%
\end{subequations}%
where $\ubar{k}_c = \mathrm{min}(k ~| ~t_{k} \geq t^{\mathrm{in}}_c),~\bar{k}_c = \mathrm{max}(k ~| ~t_{k} \leq t^{\mathrm{out}}_c),~c \in \{i,j\}$ and $t_k = k\Delta t$. $\bar{l}$ is a constant accounting for the change of coordinates between platoons $i$ and $j$, as illustrated in Figure~\ref{fig:ori-adjust} and inspired by the \emph{virtual platooning} concept \cite{MoralesMedina2017}. Note that either~\eqref{eq:mip-sdca-a} or~\eqref{eq:mip-sdca-b} is active w.r.t. the selected sequence, i.e., $\iota_k(j) < \iota_k(i)$ means $j$ comes first before $i$ and vice-versa, within certain timeslots, i.e., $[\ubar{k}_j, \bar{k}_i]~\lor~[\ubar{k}_i, \bar{k}_j]$, respectively. 

The following conditions relate the timeslots to the platoons~$i,j$ and CZ positions 
\begin{align}
    p_i(t^{\mathrm{in}}_{\ubar{k},i}) &\geq p^{\mathrm{in}},~ p_i(t^{\mathrm{out}}_{\bar{k},i}) \leq p^{\mathrm{out}}, \\
    p_j(t^{\mathrm{in}}_{\ubar{k},j}) &\geq p^{\mathrm{in}},~    p_j(t^{\mathrm{out}}_{\bar{k},j}) \leq p^{\mathrm{out}}, \\
    p_i(t^{\mathrm{in}}_{i}) &= p^{\mathrm{in}},~p_i(t^{\mathrm{out}}_{i}) = p^{\mathrm{out}}, \\
    p_j(t^{\mathrm{in}}_{j}) &= p^{\mathrm{in}},~p_j(t^{\mathrm{out}}_{j}) = p^{\mathrm{out}}.        
\end{align}
As we consider platoons as single vehicles (Assumption~\ref{ass:cav-role}), this constraint must account for the \emph{platoon length}~$l^\text{p}_{i,k}$. 

Accordingly, binary indicators $\rho^{\mathrm{in}}_{i,j,k}, \rho^{\mathrm{out}}_{i,j,k} \in \{0,1\}$ are introduced to activate the constraint within the selected timeslots, which rewrites the constraint as
\begin{subequations}
    \begin{align}
        (\rho^{\mathrm{in}}_{i,j,k} - \rho^{\mathrm{out}}_{i,j,k})(1-r_{i,j})(p_{j,k} - l^\text{p}_{j,k} - p_{i,k} - d^{\mathrm{min}} - \bar{l} ) &\geq 0, \\
        (\rho^{\mathrm{in}}_{i,j,k} - \rho^{\mathrm{out}}_{i,j,k})(r_{i,j})(p_{i,k} - l^\text{p}_{i,k} - p_{j,k} - d^{\mathrm{min}} - \bar{l}) &\geq 0,  
    \end{align}%
    \label{eq:minlp-sdca}%
\end{subequations}%
where $r_{i,j} \in \{0,1\}$ is a binary variable defining whether $i$ crosses before $j$ ($r_{i,j} = 1$) or the converse ($r_{i,j} = 0$). 

\begin{figure}[]
    \begin{center}
    \graphicspath{{./figures/}}
    \includegraphics[scale = 0.5]{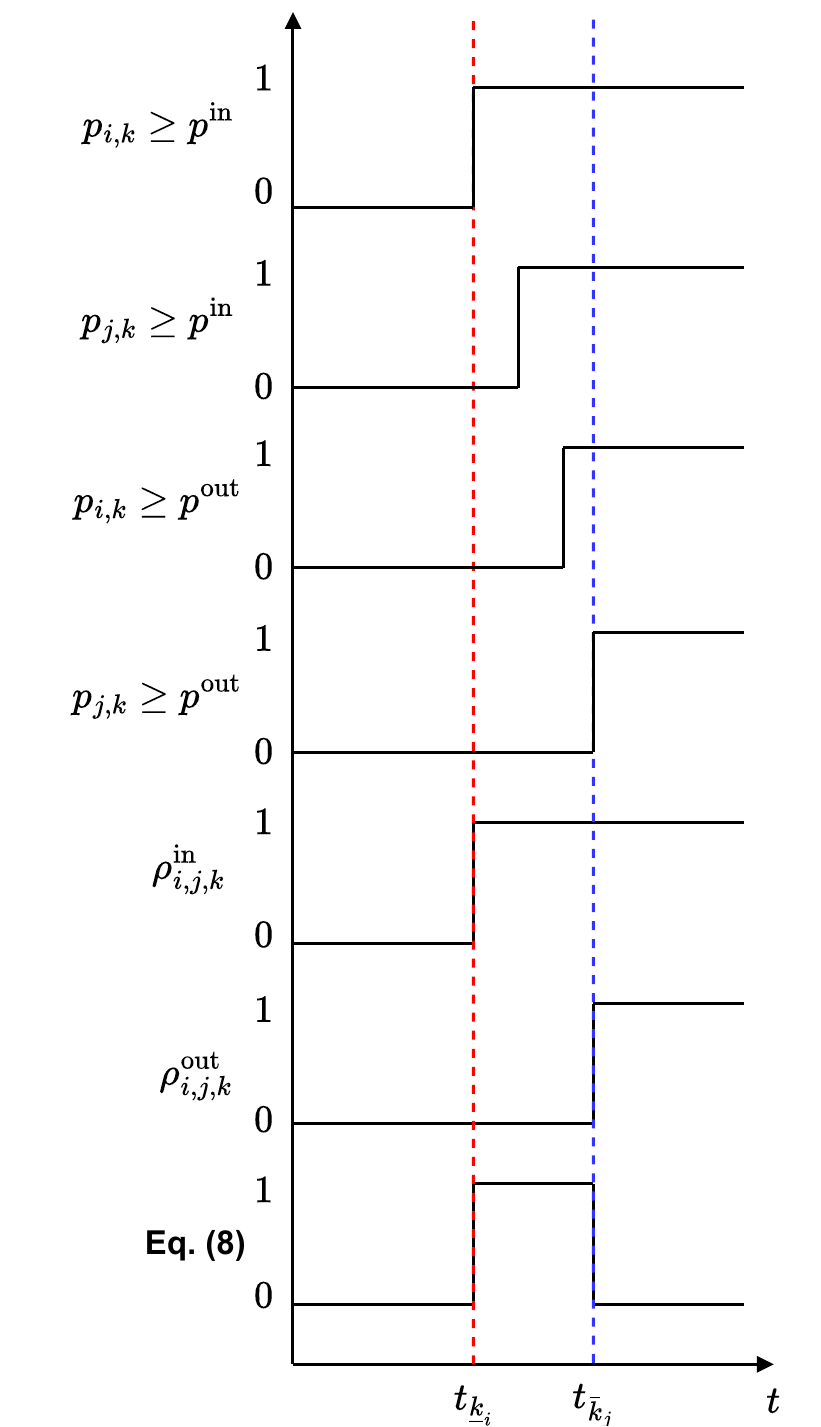}    
        \caption{Timing diagram of the safety constraint~\eqref{eq:minlp-sdca} activation, where the entry condition is based on OR logic, while the exit condition is based on AND.}
        \label{fig:timing}%
    \end{center}%
\end{figure}

The values of $\rho^{\mathrm{in}}_{i,j,k}, \rho^{\mathrm{out}}_{i,j,k}$ depend on the times at which the platoons occupy the CZ, i.e., they must satisfy these conditions
\begin{subequations}
    \label{eq:position_constraints}
    \begin{align}
        \rho^{\mathrm{in}}_{i,j,k} &\leq 1 + \frac{\bar{p}^{\mathrm{in}}_{t} - p^{\mathrm{in}}}{M^{\mathrm{b}}},  \label{eq:in-cond-0}  \\    
        \rho^{\mathrm{in}}_{i,j,k} &\geq \frac{\bar{p}^{\mathrm{in}}_{t} - p^{\mathrm{in}}}{M^{\mathrm{b}}}, \label{eq:in-cond-1}  \\ 
    \rho^{\mathrm{out}}_{i,j,k} &\leq 1 + \frac{\bar{p}^{\mathrm{out}}_{t} - p^{\mathrm{out}}}{M^{\mathrm{b}}}, \label{eq:out-cond-0}   \\        
        \rho^{\mathrm{out}}_{i,j,k} &\geq \frac{\bar{p}^{\mathrm{out}}_{t} - p^{\mathrm{out}}}{M^{\mathrm{b}}}, \label{eq:out-cond-1}     
    \end{align}
\end{subequations}
where,
\begin{align}
    \bar{p}^{\mathrm{in}}_{t} &= r_{i,j}p_{i,k} + (1-r_{i,j})p_{j,k}, \\  
    \bar{p}^{\mathrm{out}}_{t} &=  r_{i,j}p_{j,k} + (1-r_{i,j})p_{i,k},         
    \label{eq:pos-and}
\end{align}
and $M^{\mathrm{b}}$ is a sufficiently large constant value, i.e., Big-M~\cite{conforti_integer_2014}. Conditions~\eqref{eq:in-cond-0} and~\eqref{eq:in-cond-1} are used to set $\rho^{\mathrm{in}}_{i,j,k}$ to $0$ when \emph{either} of the platoons $i,j$ (OR condition) is before $p^{\mathrm{in}}$ and to $1$ otherwise. Similarly, conditions~\eqref{eq:out-cond-0} and~\eqref{eq:out-cond-1} are used to set $\rho^{\mathrm{out}}_{i,j,k} $ to $0$ or $1$, respectively when \emph{both} platoons (AND condition) are before or after $p^{\mathrm{out}}$. The implication of these conditions to constraint~\eqref{eq:minlp-sdca} is illustrated in Figure~\ref{fig:timing}. 

Avoiding the multiplication of integer variables in \eqref{eq:minlp-sdca} is convenient from a coordination problem formulation standpoint~\cite{conforti_integer_2014}, as shown next. Indeed, the conditions~\eqref{eq:minlp-sdca}-\eqref{eq:pos-and} can be rewritten as follows
\begin{subequations}
    \begin{align}
        M^{\mathrm{b}}(1-\rho^{\mathrm{in}}_{i,j,k} + \rho^{\mathrm{out}}_{i,j,k} + 1-r_{i,j}) &+ \nonumber \\ 
        p_{j,k} - l^\text{p}_{j,k} - p_{i,k} - d^{\mathrm{min}} - \bar{l} &\geq 0, \\
        M^{\mathrm{b}}(1-\rho^{\mathrm{in}}_{i,j,k} + \rho^{\mathrm{out}}_{i,j,k} + r_{i,j}) &+ \nonumber \\
        p_{i,k} - l^\text{p}_{i,k} - p_{j,k} - d^{\mathrm{min}} - \bar{l} &\geq 0,         
    \end{align}%
    \label{eq:miqp-sdca}%
\end{subequations}%
along with the position constraints~\eqref{eq:position_constraints},~\eqref{eq:pos-and} 
\begin{subequations}
    \begin{align}    
        p_{i,k} - p^{\mathrm{in}} &\leq M^{\mathrm{b}}\rho^{\mathrm{in}}_{i,j,k}, \label{eq:in-pi-1} \\
        p_{j,k} - p^{\mathrm{in}} &\leq M^{\mathrm{b}}\rho^{\mathrm{in}}_{i,j,k}, \label{eq:in-pj-1} \\          
        -2M^{\mathrm{b}}(1-r_{i,j}) - p_{i,k} + p^{\mathrm{in}} &\leq -M^{\mathrm{b}}(\rho^{\mathrm{in}}_{i,j,k}-1), \label{eq:in-pi-0} \\
        -2M^{\mathrm{b}}(r_{i,j}) - p_{j,k} + p^{\mathrm{in}} &\leq -M^{\mathrm{b}}(\rho^{\mathrm{in}}_{i,j,k}-1), \label{eq:in-pj-0}  \\
        -2M^{\mathrm{b}}(r_{i,j}) + p_{i,k} - p^{\mathrm{out}} &\leq M^{\mathrm{b}}\rho^{\mathrm{out}}_{i,j,k}, \label{eq:out-pi-1} \\
        -2M^{\mathrm{b}}(1-r_{i,j}) + p_{j,k} - p^{\mathrm{out}} &\leq M^{\mathrm{b}}\rho^{\mathrm{out}}_{i,j,k}, \label{eq:out-pj-1} \\   
        - p_{i,k} + p^{\mathrm{out}} &\leq -M^{\mathrm{b}}(\rho^{\mathrm{out}}_{i,j,k}-1), \label{eq:out-pi-0} \\
        - p_{j,k} + p^{\mathrm{out}} &\leq -M^{\mathrm{b}}(\rho^{\mathrm{out}}_{i,j,k}-1). \label{eq:out-pj-0}          
    \end{align} 
    \label{eq:sdca-ideal}    
\end{subequations}
It can be verified that~\eqref{eq:in-pi-1},~\eqref{eq:in-pj-1} imply~\eqref{eq:in-cond-1}, while~\eqref{eq:in-pi-0},~\eqref{eq:in-pj-0} correspond to~\eqref{eq:in-cond-0}. Similarly,~\eqref{eq:out-pi-0},~\eqref{eq:out-pj-0} imply~\eqref{eq:out-cond-1}, while~\eqref{eq:out-pi-0},~\eqref{eq:out-pj-0} correspond to~\eqref{eq:out-cond-0}.

\subsubsection{Rear-end collision avoidance}
Consider two adjacent vehicles coming from the same direction, with platoon $i$ behind platoon $j$. To avoid rear-end collisions, the position gap between the two platoons must be no smaller than $d^{\mathrm{min}}$, i.e.,
\begin{align}
    p_{j,k} - l^\text{p}_{j,k} - p_{i,k} \geq d^{\mathrm{min}}.
    \label{eq:reca}
\end{align}
%

%
\subsection{Objective Function}
A CAV $i \in \mathcal{N}$ aims at following its reference speed profile $v_{i,k}^{\mathrm{ref}}$ while also minimizing its acceleration/deceleration over some \emph{prediction} horizon $K^\text{p} \in \mathbb{N}$ (or $T^\text{p} \in \mathbb{R}_{+}$). This can be formalized by
\begin{align}
    J_{i} = &\sum_{k=\lfloor{\frac{t}{\Delta t}}\rfloor}^{K^\text{p}-1} q^{\mathrm{v}}(v^{\mathrm{ref}}_{i,k} - v_{i,k})^2 + q^{\mathrm{u}}u_{i,k}^{2}  + q^{\mathrm{v}}(v^{\mathrm{ref}}_{i,k} - v_{i,k})^2, 
    \label{eq:obj-vt}     
\end{align}
where $q^{\mathrm{v}}$ and $q^{\mathrm{u}}$ are constant weights. Accordingly, ~$v^{\mathrm{ref}}_{i,k}$ is defined as 
\begin{align}
    v^{\mathrm{ref}}_{i,k} = &\begin{cases}
    v_{m,k} \quad &l^{p}_{i,k-1} \geq \bar{d}, \\      
    v^{\mathrm{nom}} \quad &l^{p}_{i,k-1} < \bar{d}, 
    \end{cases}
    \label{eq:ref_speed}
\end{align}
where~$m$ is the index of the HDV following the \emph{leading} CAV~$i$ and $v^{\mathrm{nom}} \geq v_{m,k}$ is the $i$-th CAV's preferred speed. The reference speed in~\eqref{eq:ref_speed} aims at limiting the platoon length~$l^{p}_{i,k-1}$ in case the following $m$-th HDV is falling behind and is set to be constant over $K^{\mathrm{p}}$ steps for the prediction purpose.


\subsection{Problem Formulation}
The vehicle coordination problem can be formulated at the time~$t$ as the following mixed-integer quadratic programming (MIQP) constrained optimal control problem
\begin{subequations}
    \begin{align}
        \Phi^{\mathrm{OMIQP}}(\textbf{v}^{\mathrm{ref}}_{t}, \pmb{x}_{0}, \textbf{p}^\text{h}) &= \nonumber \\ 
        \min_{\textbf{r}, \boldsymbol{\rho}, \textbf{w}} \quad &\sum_{i}^{N} J_{i}(\textbf{w}_i) \label{eq:miqp-ori-obj} \\
        \text{s.t.} \quad &x_{i,k+1} = Ax_{i,k} + Bu_{i,k}, \label{eq:miqp-ori-dyn} \\
        &x_{i,0} = x^{\mathrm{0}}_{i}, \\        
       &v^{\mathrm{min}} \leq v_{i,k} \leq v^{\mathrm{max}}, \label{eq:miqp-ori-v}
        \\
        &u^{\mathrm{min}} \leq u_{i,k} \leq u^{\mathrm{max}}, \label{eq:miqp-ori-u}
        \\           
        & \text{Eq.}~\eqref{eq:miqp-sdca},~\eqref{eq:sdca-ideal},~\eqref{eq:reca}, \label{eq:miqp-ori-sdca} \\  
        &r_{i,j}, \rho^{\mathrm{in}}_{i,j,k}, \rho^{\mathrm{out}}_{i,j,k} \in \{0,1\}, \label{eq:miqp-cons-r}%
    \end{align}%
    \label{eq:miqp-ori}%
\end{subequations}%
where $\textbf{p}^\text{h} = \left[ \textbf{p}_{i,0:T^\text{p}},~...,~ \textbf{p}_{M,0:T^\text{p}} \right]^{\top}$ collects the predicted trajectories of \emph{tail} HDVs, which define the platoon length $l^\mathrm{p}_{i,k}$ appearing in~\eqref{eq:miqp-sdca}, and~\eqref{eq:reca},~$\pmb{x}_{0} = \left[ x_{1,0}... ,~ x_{N,0} \right]^{\top}$, collects the initial states of all CAVs, $\textbf{r} = \left[ r_{1,2} ,~ ... ,~ r_{i,j} ,~ ... ,~ r_{N-1,N} \right]^{\top}$, $\boldsymbol{\rho}^{\mathrm{in}} = \left[ \rho^{\mathrm{in}}_{1,2,0} ,~ ... ,~ \rho^{\mathrm{in}}_{N-1,N,K^{\mathrm{p}}}  \right]^{\top}$, $\boldsymbol{\rho}^{\mathrm{out}} = \left[ \rho^{\mathrm{out}}_{1,2,0} ,~ ... ,~ \rho^{\mathrm{out}}_{N-1,N,K^{\mathrm{p}}} \right]^{\top}$,~$\boldsymbol{\rho} = \left[ (\boldsymbol{\rho}^{\mathrm{in}})^{\top},~ (\boldsymbol{\rho}^{\mathrm{out}})^{\top} \right]^{\top}$ collects the binary variables encoding the crossing order and safety constraints timing. Further, $\textbf{v}^{\mathrm{ref}}_{t} = \left[ v^{\mathrm{ref}}_{1,k} ,~ ... ,~ v^{\mathrm{ref}}_{N,k} \right]^{\top}$ collects the reference velocities, and the continuous optimization variables are lumped in~$\textbf{w} = \left[ \textbf{w}_{1} ,~ ... ,~ \textbf{w}_{i} ,~ ... ,~ \textbf{w}_{N} \right]^{\top}$, where $\textbf{w}_i = \left[ w_{i,1} ,~ ... ,~ w_{i,k} ,~ ... ,~ w_{i,K^{\mathrm{p}}} \right]^{\top} \in \mathbf{R}^{n^{\textbf{w}} \times 1}$ with $w_{i,k} = \left[ x_{i,k} ,~ u_{i,k} \right]^{\top}$ lumping together the states and control inputs of vehicle $i$. Note that the number of platoons that are scheduled to cross after platoon $i \in \{i,N-1\}$ can be obtained as~$\sum_j r_{{i,j}}, ~\forall j \in \{i+1,N\} $, and the crossing order $\mathcal{O}_k$ can be constructed from $\textbf{r}$ directly.

When solving MIQP~\eqref{eq:miqp-ori} above, by setting $r_{i,j}$ to either $0$ or $1$, the solver selects which of the two (complementary) conditions in~\eqref{eq:miqp-sdca} will be enforced to define the optimal order. To avoid infeasibility in case the initial position difference between platoons is lower than $d^{\mathrm{min}}$, a slack variable~$\eta_{i,j,k} \geq 0$ is added to the equations
\begin{subequations}
    \begin{align}
        M^{\mathrm{b}}(1-\rho^{\mathrm{out}}_{i,j,k} + \rho^{\mathrm{in}}_{i,j,k} + 1-r_{i,j}) &+ \nonumber \\ 
        p_{j,k} - l^\text{p}_{j,k} - p_{i,k} - d^{\mathrm{min}} - \bar{l} + \eta_{i,j,k} &\geq 0, \\
        M^{\mathrm{b}}(1-\rho^{\mathrm{out}}_{i,j,k} + \rho^{\mathrm{in}}_{i,j,k} + r_{i,j}) &+ \nonumber \\
        p_{i,k} - l^\text{p}_{i,k} - p_{j,k} - d^{\mathrm{min}} - \bar{l} + \eta_{i,j,k} &\geq 0,         
    \end{align}%
    \label{eq:sdca-slack}%
\end{subequations}
and a linear term is introduced in the cost to penalize this relaxation so that Problem~\eqref{eq:miqp-ori} is reformulated as
\begin{subequations}
    \begin{align}
        \Phi^{\mathrm{OMIQP}}(\textbf{v}^{\mathrm{ref}}_{t}, \pmb{x}_{0}, \textbf{p}^\text{h}) &= \nonumber \\ 
        \min_{\textbf{r}, \boldsymbol{\rho}, \textbf{w}, \boldsymbol{\eta}} \quad &\sum_{i}^{N} J_{i}(\textbf{w}_i) + \sum_{i=1}^{N-1} \sum_{j=i+1}^{N} J^{\mathrm{e}}(\boldsymbol{\eta}_{i,j}) \label{eq:miqp-rel-obj} \\
        \text{s.t.} \quad &x_{i,k+1} = Ax_{i,k} + Bu_{i,k}, \label{eq:miqp-rel-dyn} \\
        &x_{i,0} = x^{\mathrm{0}}_{i}, \\        
       &v^{\mathrm{min}} \leq v_{i,k} \leq v^{\mathrm{max}}, \label{eq:miqp-rel-v}
        \\
        &u^{\mathrm{min}} \leq u_{i,k} \leq u^{\mathrm{max}}, \label{eq:miqp-rel-u}
        \\           
        & \text{Eq.}~\eqref{eq:sdca-slack},~\eqref{eq:sdca-ideal},~\eqref{eq:reca}, \label{eq:miqp-rel-sdca} \\  
        &r_{i,j}, \rho^{\mathrm{in}}_{i,j,k}, \rho^{\mathrm{out}}_{i,j,k} \in \{0,1\}, \label{eq:miqp-cons-r}        
    \end{align}
    \label{eq:miqp-relax}%
\end{subequations}
where,
\begin{align}
    J^{\mathrm{e}} = \sum_{k=\lfloor{\frac{t}{\Delta t}}\rfloor}^{K^\text{p}-1} q^{\mathrm{e,l}}\eta_{i,j,k} + \sum_{k=\lfloor{\frac{t}{\Delta t}}\rfloor}^{K^\text{p}-1} q^{\mathrm{e,q}}(\eta_{i,j,k})^2.
    \label{eq:obj-slack}
\end{align}
Variable $\boldsymbol{\eta}_{i,j} = \begin{bmatrix} \eta_{i,j,0} & ... & \eta_{i,j,K^{\mathrm{p}}}  \end{bmatrix}$ collects the slacks for each pair $i \in \{1,N-1\},j \in \{i+1,N\},~ i \neq j$ for all $k$, and $q^{\mathrm{e,l}}, q^{\mathrm{e,q}}$ are fixed weights for the slack variable penalty cost~\eqref{eq:obj-slack}, which consists of linear and quadratic terms. The quadratic term is in principle not needed, but we add it to the cost to introduce some positive curvature that can help the solver converge faster. From this point on, we denote the MIQP~\eqref{eq:miqp-relax} as the original MIQP (OMIQP).

The MIQP is executed in a closed-loop fashion according to Algorithm~\ref{alg:MIQP}, lines 4-5. With respect to a standard closed-loop implementation, in this case, we need to account for the fact that vehicles will eventually reach the intersection's entry, such that the definition of an order will eventually become meaningless. To address that issue, we apply a simple strategy of not updating the crossing order once the vehicles are too close to the intersection. Accordingly, the fixed-order counterpart of MIQP~\eqref{eq:miqp-relax}, given in~\eqref{eq:miqp-fo}, is solved on lines 7-8 to keep yielding CAVs control input.

In the context of continuous traffic flow, our algorithm can be readily modified to take into account the incoming platoons and subsequently coordinate their order and trajectories. Once a platoon reaches the entry, we can fix its sequence and exclude it from the MIQP problem. For simplicity, we do not consider this scheme for now and instead use a static number of platoons. 

\algrenewcommand \algorithmicrequire{\textbf{Input:}}
\algrenewcommand \algorithmicensure{\textbf{Output:}}

\begin{algorithm}[]
\begin{algorithmic}[1]
\caption{MIQP} \label{alg:MIQP}
\Require $\textbf{v}^{\mathrm{ref}}_{t}, \pmb{x}_{0}, \textbf{p}^{\mathrm{h}}$
\Ensure $\mathcal{O}_{t}$, \textbf{w}
\For{$t \in \mathbb{R}^{+}$}  
        \State Obtain $\textbf{v}^{\mathrm{ref}}_{t}, \pmb{x}_{0}, \textbf{p}^{\mathrm{h}}$ \Comment{Parameter}
    \If{$p_{i,k} \leq p^{\mathrm{in}}, ~\forall i \in \mathcal{M}, \mathcal{N}$}           
        \State Solve MIQP $\Phi^{\mathrm{U}}$~\eqref{eq:miqp-relax}~/~\eqref{eq:miqp-simple} 
        \State Obtain $\mathcal{O}_k$ from $\textbf{r}$
    \Else
        \State Set $\mathcal{O}_k = \mathcal{O}_{t-1}$ and obtain $\textbf{r}$
        \State Solve fixed-order QP~\eqref{eq:miqp-fo} 
    \EndIf        
    \State Apply $\textbf{w}$ to CAVs of platoons     
\EndFor        
\end{algorithmic}
\end{algorithm}

\subsection{Lower-complexity/Simplified MIQP}
Solving the ideal/original MIQP~\eqref{eq:miqp-relax} which imposes the safety constraints~\eqref{eq:sdca-slack} in a closed-loop way can be computationally heavy as the timing binaries $\boldsymbol{\rho}$ are strictly upper- and lower-bounded at each time $k$, as shown by conditions~\eqref{eq:sdca-ideal}. However, the mechanism illustrated in Figure~\ref{fig:timing} can still be realized without~\eqref{eq:in-pi-0}-\eqref{eq:out-pj-1}. This is because they are used to deactivate constraint~\eqref{eq:sdca-slack}, i.e., by setting $\rho^{in}_{i,j,k}=0$ \& $\rho^{out}_{i,j,k}=1$, which implies less restriction on the solution space and potentially produces solutions with lower costs. Therefore, the solver will try to achieve it without the presence of~\eqref{eq:in-pi-0}-\eqref{eq:out-pj-1} anyway. 

Accordingly, we can propose the following simplified conditions
\begin{subequations}
    \begin{align}
        p_{i,k} - p^{\mathrm{in}} &\leq M^{\mathrm{b}}\rho^{\mathrm{in}}_{i,j,k}, \\
        p_{j,k} - p^{\mathrm{in}} &\leq M^{\mathrm{b}}\rho^{\mathrm{in}}_{i,j,k}, \\           
        -p_{i,k} + p^{\mathrm{out}} &\leq -M^{\mathrm{b}}(\rho^{\mathrm{out}}_{i,j,k} - 1), \label{eq:sim-out-pi} \\    
        -p_{j,k} + p^{\mathrm{out}} &\leq -M^{\mathrm{b}}(\rho^{\mathrm{out}}_{i,j,k} - 1), \label{eq:sim-out-pj}  \\       
        \rho^{\mathrm{in}}_{i,j,k} &\leq \rho^{\mathrm{in}}_{i,j,k+1}, \label{eq:pi-lead} \\
        \rho^{\mathrm{out}}_{i,j,k} &\leq \rho^{\mathrm{out}}_{i,j,k+1}. \label{eq:pj-lead}        
    \end{align}
    \label{eq:sdca-simple}
\end{subequations}
Additionally, we introduce less complex conditions~\eqref{eq:pi-lead}-\eqref{eq:pj-lead} to prevent activation of~\eqref{eq:sdca-slack} at time $k$ before $k+1$ is active by exploiting the fact that $v^{\mathrm{min}}>0$, i.e., vehicles are closer to the intersection in each time $t$. 

As we will demonstrate in the simulations part, i.e., Section~\ref{sec:simulation}, this simplification yields an MIQP that can be solved much faster than the original one and converges within the imposed solver iterations limit. The simplified MIQP (SMIQP) is formulated as follows
\begin{subequations}
    \begin{align}
        \Phi^{\mathrm{SMIQP}}(\textbf{v}^{\mathrm{ref}}_{t}, \pmb{x}_{0}, \textbf{p}^\text{h}) &= \nonumber \\ 
         \min_{\textbf{r}, \boldsymbol{\rho}, \textbf{w}, \boldsymbol{\eta}} \quad &\sum_{i}^{N} J_{i}(\textbf{w}_i) + \sum_{i=1}^{N-1}\sum_{j=2}^{N} J^{\mathrm{e}}(\boldsymbol{\eta}_{i,j}) \label{eq:miqp-sim-obj} \\
        \text{s.t.} \quad &x_{i,k+1} = Ax_{i,k} + Bu_{i,k}, \label{eq:miqp-sim-dyn} \\
        &x_{i,0} = x^{\mathrm{0}}_{i}, \\        
       &v^{\mathrm{min}} \leq v_{i,k} \leq v^{\mathrm{max}}, \label{eq:miqp-sim-v}
        \\
        &u^{\mathrm{min}} \leq u_{i,k} \leq u^{\mathrm{max}}, \label{eq:miqp-sim-u}
        \\           
        & \text{Eq.}~\eqref{eq:sdca-slack},~\eqref{eq:sdca-simple},~\eqref{eq:reca}, \label{eq:miqp-sim-sdca} \\ 
        &r_{i,j}, \rho^{\mathrm{in}}_{i,j,k}, \rho^{\mathrm{out}}_{i,j,k} \in \{0,1\}. \label{eq:miqp-cons-s}        
    \end{align}
    \label{eq:miqp-simple}
\end{subequations}
As this problem is an alternative to MIQP~\eqref{eq:miqp-relax}, we will also adopt the closed-loop strategy described in Algorithm~\ref{alg:MIQP}.

%
\section{Heuristic Approach}
\label{sec:heuristic}
As mentioned before, we aim to develop a heuristic algorithm to avoid solving MIQP~\eqref{eq:miqp-relax} or~\eqref{eq:miqp-simple} for closed-loop applications. The main motivation is to mitigate the computational burden associated with solving MIQPs, as this can hinder real-time applicability.

The heuristic we propose exploits the structure of the problem. In particular, our heuristic is built on the fact that if the current crossing order $\mathcal{O}_k$ is leading to an accident or no longer beneficial, a safe (feasible) coordination can be obtained by simply switching to a better alternative order.

The basic idea is as follows. At every time step, we monitor the safety constraints to check whether they will be violated or not in the future. If any platoon consistently violates the constraint, we try to swap its order with an adjacent platoon; we compare the cost of the two (current and alternative) orders; and we retain the order with the lowest cost. To evaluate multiple alternative orders, a depth-first branching strategy~\cite{conforti_integer_2014} as illustrated in Figure~\ref{fig:bnb-diagram} is applied for the cost comparison.

The complete pseudocode for the proposed heuristic is given in Algorithm~\ref{alg:heuristic}, which we explain in detail next.

\subsection{One-Time MIQP}

During initialization, we obtain the first crossing order $\mathcal{O}_{t=0}$ by solving the MIQP~\eqref{eq:miqp-simple}, which is then designated as the \emph{initial} current order $\mathcal{O}_0$, see line 3 of Algorithm~\ref{alg:heuristic}. The MIQP is solved only once in the beginning so that the order used by the platoons to start with is guaranteed to be both feasible and optimal. Also, the resulting position trajectories are used to approximate timing parameter $\hat{\boldsymbol{\rho}}$, which we then use to solve the QP~\eqref{eq:miqp-fo} for the next time $t$. This will be discussed in detail in cost comparison part~\ref{subsec:cost_comparison}.

As an alternative to solving the MIQP, one may opt for heuristic methods, such as, e.g., First-Come-First-Serve (FCFS) or predicted arrival time to intersection, to reduce the computational burden. Future research will aim at finding alternative approaches to address this problem.


\subsection{Consistency Check}

A change of order between a pair of adjacent platoons may be necessary when the position gap between them becomes smaller. To that end, we perform the following consistency check:

\paragraph{Step 1} at each time step $t>0$, we check the collision avoidance constraints~\eqref{eq:miqp-sdca}-\eqref{eq:sdca-ideal} over the prediction horizon $T^\text{p}$ for each pair of platoons $i,j$ (see lines 8-16 in Algorithm~\ref{alg:heuristic}). In particular, this check is applied to any pair whose \emph{preceding} one is (currently) a \emph{mixed} platoon as it contains HDV(s) which may considerably change their trajectories. Hence, the following platoons are collected in the set $\mathcal{N}^{T}_t$ (lines 4 and 23).

To perform the check we construct the vector
\begin{align}
    h_i^{\mathrm{s}} &= 
    \begin{bmatrix} 
    d^{\mathrm{min}} - p_{j,0} + l^{\mathrm{p}}_{j,0} + p_{i,0} + \bar{l}, \\
                        . \\
                        . \\
    d^{\mathrm{min}} - p_{j,k} + l^{\mathrm{p}}_{j,k} + p_{i,k} + \bar{l}, \\                                      . \\
                        . \\
    d^{\mathrm{min}} - p_{j,T^\text{p}} + l^{\mathrm{p}}_{j,K^\mathrm{p}} + p_{i,K^\mathrm{p}} + \bar{l},    
    \end{bmatrix},  ~\forall i \in \mathcal{N}^{\mathrm{T}}_t,
    \label{eq:safety-cons}   
\end{align}
where $l^{\mathrm{p}}_{j,k}$ are computed from predicted HDV trajectories using~\eqref{eq:hdv-cd} (line 6) and the optimal position of the CAVs $p_{i,k}, p_{j,k}$ are obtained from the solution of~\eqref{eq:miqp-fo} at the previous time $t-1$ (line 9). 

\paragraph{Step 2} A (predicted) violation is obtained if any of the components of $h_i^{\mathrm{s}}$ becomes positive. We keep track of how many constraint violations occur by the scalar $s_i$ (line 11). We use this variable to trigger a potential reorder whenever $s_i=n^\text{max}$, where $n^\text{max}$ is a fixed parameter (line 13).

Setting $n^\text{max}>1$ allows us to robustify against false positive triggers, i.e., a reordering that is then followed by another reordering that restores the original order, which might be due to, e.g., noise, and which can cause chattering behavior.

\paragraph{Step 3} Whenever $s_i=n^{\mathrm{max}}$ for some $i$, we add $i$ to set $\mathcal{E}_t$ (lines 13 - 15). This set will be used next to restrict the reordering procedure to only consider potential swaps between platoon $i$ and the preceding one $j$, i.e., only $i\in\mathcal{E}_t$ will be considered in the cost comparison step explained next.

\subsection{Cost Comparison}
\label{subsec:cost_comparison}

As anticipated above, the reordering procedure is guided by the set $\mathcal{E}_t$, such that it only targets specific sets of platoons that have been consistently violating~\eqref{eq:safety-cons}, see line 17 in Algorithm~\ref{alg:heuristic}.

For each platoon $i\in\mathcal{E}_t$, the algorithm performs cost comparisons of two subproblems of the current and alternative (swapped) order given the current measurements/parameter at time $t$.

\begin{enumerate}
    \item \textbf{Subproblem A}: Solve OCP~\eqref{eq:miqp-relax}~/~\eqref{eq:miqp-simple} with the fixed, \emph{current} order, i.e., platoon $i$ follows the platoon $j$. We introduce the order with this fixed sequence as $O^{i|j}_{t-1}$, with the superscript notation here indicating the specific sequences of $i,j$. The rest of the sequences in $O^{i|j}_{t-1}$ are copied from the previous crossing order and left unchanged. 
    
    As the order is fixed, the binaries $\textbf{r}^{i|j}$ are also fixed and become parameters. To further reduce computational complexity, the timing binaries $\hat{\boldsymbol{\rho}}$ are approximated by using the predicted trajectory $\textbf{w}$ from the previous time $t-1$ (lines 3 and 29). 

    Since all the binary decision variables have become parameters, we obtain the following fixed-order QP (FO-QP) 
    \begin{subequations}
        \begin{align}
            \Phi^{\mathrm{FOQP}}_{i|j}(\textbf{v}^{\mathrm{ref}}_{t}, \pmb{x}_{0}, \textbf{p}^\text{h}, &\textbf{r}^{i|j}, \hat{\boldsymbol{\rho}}) = \\ \min_{\textbf{w}, \boldsymbol{\eta}} \quad &\sum_{i}^{N} J_{i}(\textbf{w}_i) + \sum_{i=1}^{N-1}\sum_{j=2}^{N} J^{\mathrm{e}}(\boldsymbol{\eta}_{i,j}) \label{eq:miqp-fo-obj} \\
               \text{s.t.} \quad &x_{i,k+1} = Ax_{i,k} + Bu_{i,k}, \label{eq:miqp-fo-dyn} \\
            &x_{i,0} = x^{\mathrm{0}}_{i}, \label{eq:miqp-fo-init} \\        
            &v^{\mathrm{min}} \leq v_{i,k} \leq v^{\mathrm{max}}, \label{eq:miqp-fo-v}
            \\
            &u^{\mathrm{min}} \leq u_{i,k} \leq u^{\mathrm{max}}, \label{eq:miqp-fo-u}
            \\           
            & \text{Eq.}~\eqref{eq:sdca-slack},~\eqref{eq:reca} \label{eq:miqp-fo-sdca}. 
        \end{align}
        \label{eq:miqp-fo}%
    \end{subequations}

    \item \textbf{Subproblem B}: The sequence of the pair $i,j$  is \emph{reversed} in $\mathcal{O}_k$, such that CAV $j$ follows CAV $i$. This defines an \emph{alternative} order $O^{j|i}_{t-1}$. Using this new order, we solve FO-QP~\eqref{eq:miqp-fo} $\Phi^{\mathrm{F}}_{j|i}(\textbf{v}^{\mathrm{ref}}_{t},~\pmb{x}_{0},~\textbf{p}^\text{h},~\textbf{r}^{j|i},\hat{\boldsymbol{\rho}})$, where the difference here is that we replace $\textbf{r}^{i|j}$ with $\textbf{r}^{j|i}$ to account for the different order. Note that the same timing binaries as in Subproblem A are applied here. Consequently, the lateral collision avoidance constraint will only be enforced approximately. In practice, this does not create safety concerns, as large safety margins need to be introduced anyway by allowing the change of order only before the platoons are close to the intersection; Also, the cost estimate is only approximate.
\end{enumerate}

\begin{figure}[]
    \begin{center}
    \graphicspath{{./figures/}}
    \includegraphics[scale = 0.55]{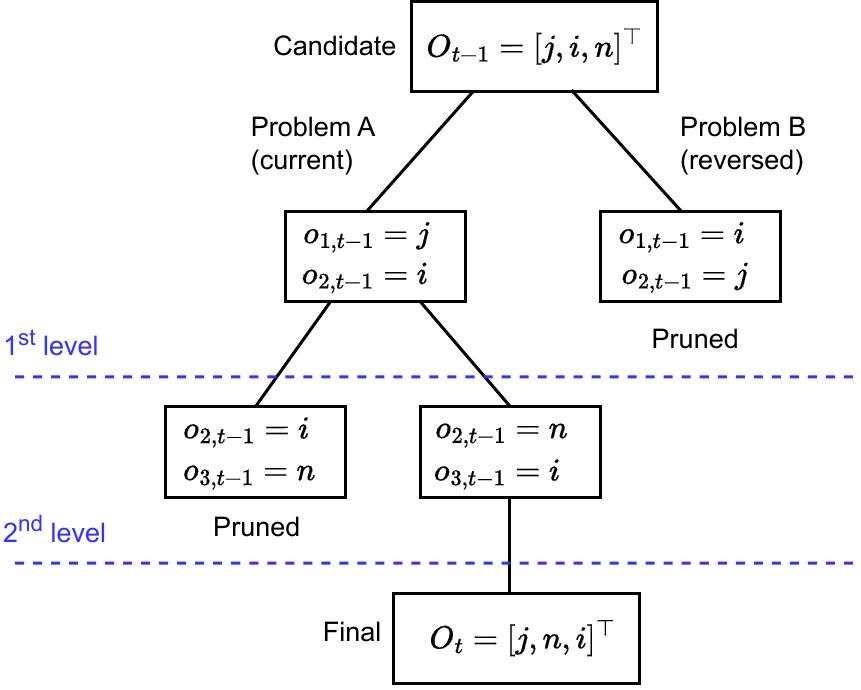}    
        \caption{Branching process of the current and alternative (reversed) subproblems performed in the heuristic}
        \label{fig:bnb-diagram}
    \end{center}
\end{figure}

Once both problems are solved, we compare their costs on lines $18-19$. If there exists more than one platoon in $\mathcal{E}_t$, a depth-first sequential branching for the cost comparison here is performed, as illustrated in Figure~\ref{fig:bnb-diagram}. Thus the next \emph{current} crossing order might be updated at each iteration (level) of the branching (line 20). Finally, the new order $\mathcal{O}_k$ will be the one associated with the subproblem yielding the lowest cost. 

Note that, if any change of order (line 22) takes place at time $t$, the set that contains the platoons that immediately comes after the mixed-platoons $\mathcal{N}^{\mathrm{T}}_t$ needs to be updated (lines 23). 

Furthermore, as for the MIQP Algorithm~\ref{alg:MIQP}, the order is allowed to change as long as no platoon is close to the CZ (line 7). The current order is kept when the order is frozen or $\mathcal{E}_t$ is empty. In case $\mathcal{E}_t$ is empty, i.e., no cost comparison takes place at time $t$, then~\eqref{eq:miqp-fo} is solved to generate CAVS control input (lines $26-27$).

%

%
\algrenewcommand \algorithmicrequire{\textbf{Input:}}
\algrenewcommand \algorithmicensure{\textbf{Output:}}

\begin{algorithm}[]
\caption{Heuristic}\label{alg:heuristic}
\begin{algorithmic}[1]
\Require $\textbf{v}^{\mathrm{ref}}_{t}, \pmb{x}_{0}, \textbf{p}^\text{h}, \textbf{r}, \hat{\boldsymbol{\rho}}$
\Ensure $\mathcal{O}_{t}, \textbf{w}$
\State Obtain $\pmb{x}_{0}, \textbf{p}^{\mathrm{h}}$  \Comment{Initial states and prediction}
\State Solve~\emph{one-time} MIQP~\eqref{eq:miqp-simple} $\Phi^{\mathrm{SMIQP}}$
\State Obtain $\hat{\boldsymbol{\rho}}$ and $\mathcal{O}_{0}$ from $\textbf{r}$ 
\State Initialize $\mathcal{N}^{\mathrm{T}}_0$ \Comment{For CAVs behind mixed-platoons}
\For{$t \in \mathbb{R}^{+}$}
    \State Obtain $\pmb{x}_{0}, \textbf{p}^{\mathrm{h}}$ \Comment{Current states and prediction} 
    \If{$p_{i,k} \leq p^{\mathrm{in}}, ~\forall i \in \mathcal{M}, \mathcal{N}$}       
        \For{$ i \in \mathcal{N}^{\mathrm{T}}_t$} 
                \State Retrieve $\textbf{w}$ from $t-1$ and compute $h_i^{\mathrm{s}}$ 
            \If{$\forall k: \exists ~h_i^{\mathrm{s}} > 0$} \Comment{Any cons. violation}
                    \State $s_i = s_i + 1$          \Comment{Violation counter}
            \EndIf
            \If{$s_i == n^{\mathrm{max}}$} \Comment{Consistency check}
                \State Set $\mathcal{E}_t \gets i$  \Comment{Set of violating vehicles}       
              \EndIf
        \EndFor    
        \For{$i \in \mathcal{E}_t$}
            \State Obtain the alternative order
            \State Solve and compare $\hat{\Phi}^{\mathrm{FOQP}}_{i|j}$ and $\hat{\Phi}^{\mathrm{FOQP}}_{j|i}$  
            \State Update $\mathcal{O}_{t+1}$ based on the cost  
        \EndFor
        \If{$\mathcal{O}_{t+1} \neq \mathcal{O}_{t}$} \Comment{Any change of order}
            \State Update $\mathcal{N}^{\mathrm{T}}_{t+1}$
        \EndIf
    \Else
        \State Set $\mathcal{O}_k = \mathcal{O}_{t-1}$ and obtain $\textbf{r}$ 
        \State Solve FO-QP~\eqref{eq:miqp-fo} 
    \EndIf        
    \State Approximate $\hat{\boldsymbol{\rho}}$         
    \State Apply \textbf{w} to CAVs of platoons
\EndFor
\end{algorithmic}
\end{algorithm}

%
\section{Numerical Simulations \& Evaluation}
\label{sec:simulation}

In this section, we perform numerical simulations to evaluate the performance of the proposed heuristic (H) and compare it to the performance of the original MIQP (OMIQP)~\eqref{eq:miqp-relax} and the simplified MIQP (SMIQP)~\eqref{eq:miqp-simple}.

In all simulations, we consider a static number of vehicles approaching a single symmetric four-junction intersection for the sake of simplicity. Two different scenarios are compared: the first one is the \emph{nominal} case, where all leading vehicles are CAVs, schematized in Figure~\ref{fig:case-1}; the second one, schematized in Figure~\ref{fig:case-2}, is the \emph{disturbance} case, where one of the leading vehicles is an HDV.

In the nominal case, all HDV trajectories can be partially regulated by the leading CAVs, which are able to slow them if necessary. In the disturbance case, the simulations are subject to noise stemming from uncertain human behavior, as the CAVs do not have any influence over the leading HDV. This allows us to evaluate the performance of our heuristic in realistic and non-ideal situations.

For each of the two cases above, we perform two different simulations. In the first one, we compare the heuristics with the MIQPs in terms of evaluation metrics, e.g., closed-loop cost, computation times, etc. as discussed in subsection~\ref{subsec:eval-metrics}, executed once due to their high computational burden. In the second one, we simulate the heuristic ten times with different HDV input bounds to evaluate how consistent their performance can be. All simulations are subject to HDVs with additive normally distributed input noise $\Delta u_k$ with average $\mu = 0$ and standard deviation $\sigma = 0.1$. The sampling time $\Delta t=0.1$ s, while the duration of the simulations is $T^{\mathrm{sim}}=8 $\text{ s}$ / K^{\mathrm{sim}}=80$ steps and $10$ s ($100$ steps) for the nominal and disturbance cases, respectively. The simulation takes longer for the latter, due to the presence of the additional leading HDV. 

Furthermore, we set the constraint relaxation weights $q^{\mathrm{e,q}} = 1, q^{\mathrm{e,l}} = 10^3$ when the order is fixed. To increase coordination safety, we introduce enlarged positions of the intersection within the vicinity of the CZ where the lateral collision avoidance constraints~\eqref{eq:sdca-slack} are active, i.e., the constraint is applied between $p^{\text{in}} - \delta^{\mathrm{in}}$ and $p^{\text{out}} + \delta^{\mathrm{out}}$. The values of the safety margins $\delta^{\mathrm{in}},~\delta^{\mathrm{out}}$ and all remaining constants are given in Table~\ref{tab:param}. 

All simulations are carried out using MATLAB with the CasADi framework~\cite{Andersson2012} on a laptop with an Intel Core-i5 processor and 16 GB of RAM. BONMIN \cite{Bonami2011} is used to solve the MIQPs, while the continuous-relaxed QP problem and the fixed-order parametric versions (Subproblems A/B) in the heuristic are solved by IPOPT~\cite{wachter_implementation_2006}. 

In these experiments, we utilize the fact that the HDV model prediction~\eqref{eq:hdv-cd} is not exact~\cite{faris_optimization-based_2022}; To account for this notion, we define a different simulation model that switches between car-following and reference velocity tracking as follows
\begin{align}
    u_{m,k} =
    \begin{cases}
        u_a \quad &\Delta p_{m,k} \geq \bar{d}, \quad \forall m \in \mathcal{M},\\      
        u_b \quad &\Delta p_{m,k} < \bar{d}, 
    \end{cases} 
\end{align}
\label{eq:hdv-switch}%
with
\begin{subequations}
    \begin{align}
        u_a &= k^{\mathrm{v}}(v_{m,k}-v^{\mathrm{ref}}_{m,k}) + \Delta u_k ,  \\
        u_b &= k^{\mathrm{p}}(\Delta p_{m,k} - d^{\mathrm{ref}}) + k^{\mathrm{d}}(v_{m,k}-v_{i,k}) + \Delta u_k , 
    \end{align}    
\end{subequations}
where $k^{\mathrm{v}}, k^{\mathrm{p}}, k^{\mathrm{d}}$ are weighting gains and we define~$\Delta p_{m,k} := p_{i,k}-p_{m,k}$. Finally, $p_{i,k},~v_{i,k}$ are the position and velocity of the vehicle immediately in front of the HDV $m$, and $\Delta u_k$ is an additive noise that represents uncertainty in the human driver's behavior.

%
\subsection{Evaluation Metrics}
\label{subsec:eval-metrics}

The following metrics are introduced to evaluate the closed-loop performance of the aforementioned methods, which are presented in Figures~\ref{fig:e1-oj}-\ref{fig:e2-pv} and Tables~\ref{tab:case-1-a}-\ref{tab:case-2-b}. 

\begin{itemize}
 
\item \emph{Crossing order} We monitor the crossing order $\mathcal{O}_k$ and its evolution over the simulation time $T^{\mathrm{sim}}$.

\item \emph{Cardinality and timing of reordering} We record the times at which the crossing order is changed (reordering/switching), i.e., $\mathcal{O}_{t} \neq \mathcal{O}_{t-1}$ are recorded in vector $\boldsymbol{\tau}^{\mathrm{OMIQP}} = [ \tau_1^{\mathrm{OMIQP}}, \tau_2^{\mathrm{OMIQP}}, ... ],~\boldsymbol{\tau}^{\mathrm{SMIQP}},~\boldsymbol{\tau}^{\mathrm{H}}$ for the OMIQP, SMIQP, and heuristic, respectively. Their respective cardinality is expressed as $|\boldsymbol{\tau}|$.

\item \emph{Closed-loop cost} 
We record the closed-loop total objective function (cost) values from all $N$ CAVs over the simulation duration ${T^\mathrm{sim}}$ s or in ${K^\mathrm{sim}}$ steps
\begin{align}
    \Phi^\mathrm{cl} := &\sum_{k=0}^{K^\mathrm{sim}} \sum_{i=1}^{N} q^{\mathrm{v}}(v^{\mathrm{ref}}_{i,k} - v_{i,k})^2 + \\ &q^{\mathrm{u}}u_{i,k}^{2} + \sum_{i=1}^{N-1}\sum_{j=i+1}^{N} J^{\mathrm{e}}(\eta_{i,j,k}) \nonumber
\end{align}
Additionally, we also present the closed-loop cost without the slack terms from the cost as $\Phi^\mathrm{cl,S\&I}$ to evaluate the cost of applied input and state reference tracking (S\&I). 

\item \emph{Maximum constraint violation} 
We record the worst-case slack from each pair of platoons used to relax the constraint~\eqref{eq:sdca-slack} within the safety margins $p^{\text{in}} - \delta^{\mathrm{in}}$ and $p^{\text{out}} + \delta^{\mathrm{out}}$ over $T^{\mathrm{sim}}$
\begin{subequations}
\begin{align}
    \boldsymbol{\eta}^{\mathrm{max}} := \max_{k,j,i} \ &\eta_{i,j,k} \\
    \mathrm{s.t.} \ & k\in \{0,\ldots,K^\mathrm{sim}\}, \\
    &i,j \in \{1,\ldots,N\}^2, \\
    &i\neq j.  
\end{align}
\end{subequations}
Note that this only accounts for the \emph{current} slack and not the future violation.

\item \emph{RMS of acceleration}  
We record the root mean square (RMS) of the acceleration \& deceleration of the CAVs, i.e., control input. This metric shows the average amount of control actions injected into each CAV at each time $t_k$
\begin{align}
     \textbf{u}^{\mathrm{RMS}} := \sqrt{\frac{1}{N \times K^{\mathrm{sim}}} \sum_{i=1}^{N} \sum_{k=0}^{K^{\mathrm{sim}}} u_{i,k}^2} 
\end{align}
\item \emph{Computation time}  
We record the worst-case computation time, over $T^{\mathrm{sim}}$ in seconds (s) required for solving the crossing order problem at a single time step $t_k$, i.e.,
\begin{align}
    \textbf{t}^{\mathrm{max}} := \max_{t_k} ~\textbf{t}^{\mathrm{comp}},
\end{align}
where $\textbf{t}^{\mathrm{comp}}$ is a vector collecting the computation times from a single simulation. 

\end{itemize}

\begin{table}[]
\centering
\caption{Constant values}
\begin{tabular}{|l|l|l|l|}
\hline
\textbf{Constant} & \textbf{Value}  & \textbf{Constant} & \textbf{Value}  \\ \hline
$T^{\mathrm{p}}$ / $K^{\mathrm{p}}$            & 2.6 or 3.5 s / 26 or 35     & $d^\text{ref}$          & $9$ m          \\ \hline
$q^\text{v}$            & $10$    & $\bar{l}$    & $2$ m          \\ \hline
$d^\text{min}$     & $4$ m   & $v^\text{nom}$          & $40$ km/h            \\ \hline
$q^{\mathrm{u}}$                & $1$    & $n^\text{max}$          & 3           \\ \hline
$q^\text{e,l}$                & $10$ or $1000$     & $k^\text{v}$          & 1            \\ \hline
$M^\text{b}$                & $10^3$   & $k^\text{p}$          & 2             \\ \hline
$\Delta t$       & $0.1$ s &  $k^\text{d}$          & 1          \\ \hline
$T^\text{sim}$ / $K^\text{sim}$    & $8$ or $10$ s / 80 or 100 & $\bar{d}$          & 7 m            \\ \hline
$v^\text{min/max}$ & $3.6 - 70$ km/h  &  $p^\text{in/out}$          & $\pm 2$    \\ \hline
$u^\text{min/max}$ & $\pm 3$ m/s$^{2}$ &  $\delta^{\mathrm{in}}$          & 13 m   \\ \hline
$q^\text{e,q}$ & $1$ &  $\delta^{\mathrm{out}}$          & 8 m   \\ \hline
\end{tabular}    
\label{tab:param}
\end{table}

%
\subsection{Nominal Scenario}
\label{subsec:exp-1}

\begin{figure}[t]
    \begin{center}
    \graphicspath{{./figures/}}
    \includegraphics[scale = 0.5]{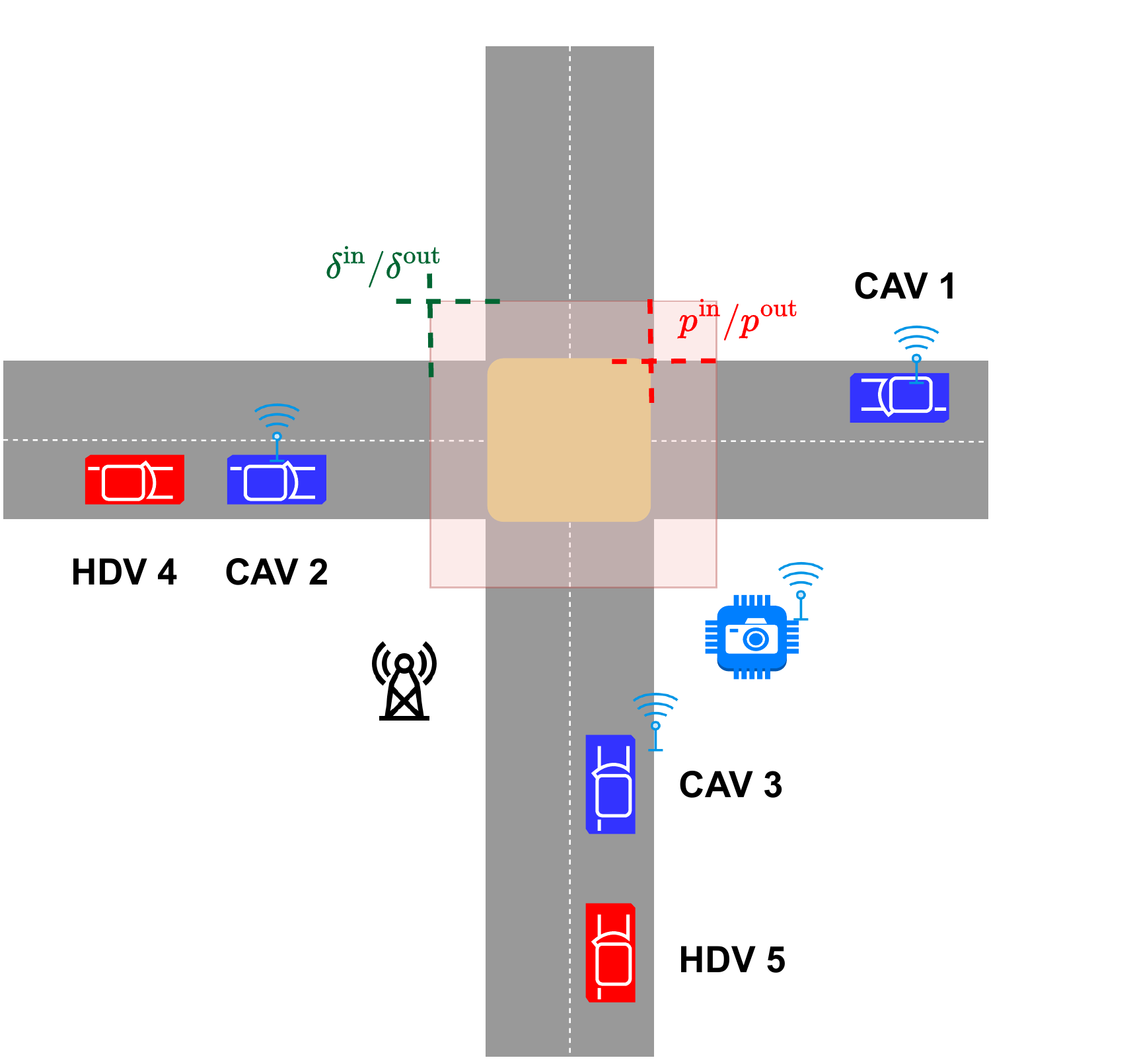}    
        \caption{Vehicle configuration for \emph{nominal} case}
        \label{fig:case-1}
    \end{center}
\end{figure}
We consider now the case in which all the leading vehicles are CAVs, as shown in Figure~\ref{fig:case-1}, where one can see that the three CAVs are colored blue and the two HDVs are colored red. Platoon 2 is composed of CAV 2 and HDV 4; Platoon 3 is composed of CAV 3 and HDV 5; and CAV 1 is a one-vehicle platoon. 

The first vehicle on the line is CAV 2, which starts at a distance of $53$ m from the center of CZ. The rest of the vehicles in the line, with the sequence of HDV 4 - CAV 3 - HDV 5 - CAV 1, are placed with a gap of 7.5 m between each other. The initial speed of all vehicles is $v^{0}_{i} = 50 \text{ km/h } (13.89 $ m/s), and the nominal reference speed is $v^\text{nom} = 60$ km/h ($16.67$ m/s). In order to force a switch in the crossing order, HDV 4 is set to slow down to $23$ km/h ($6.39$ m/s). The reordering problem is solved in a closed loop until CAV 1 has reached the CZ, and the order is fixed after that time.

\subsubsection{Simulation against MIQPs}
\label{subsubsec:exp-1-a}

\begin{figure*}[]
    \begin{center}
        \begin{minipage}{.5\textwidth}
            \centering
            \includegraphics[width=\linewidth]{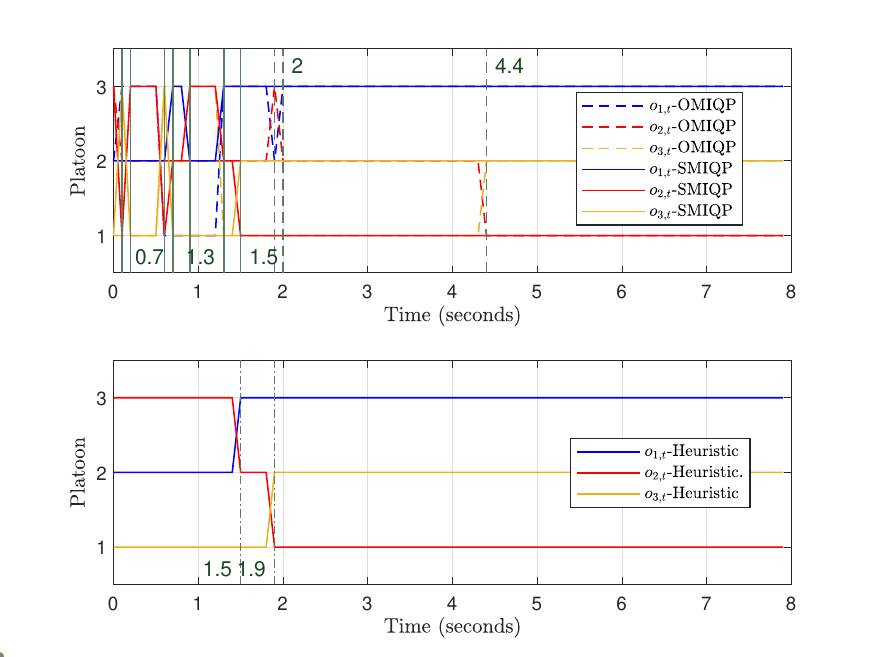}
        \end{minipage}%
        \begin{minipage}{.5\textwidth}
            \centering
            \includegraphics[width=\linewidth]{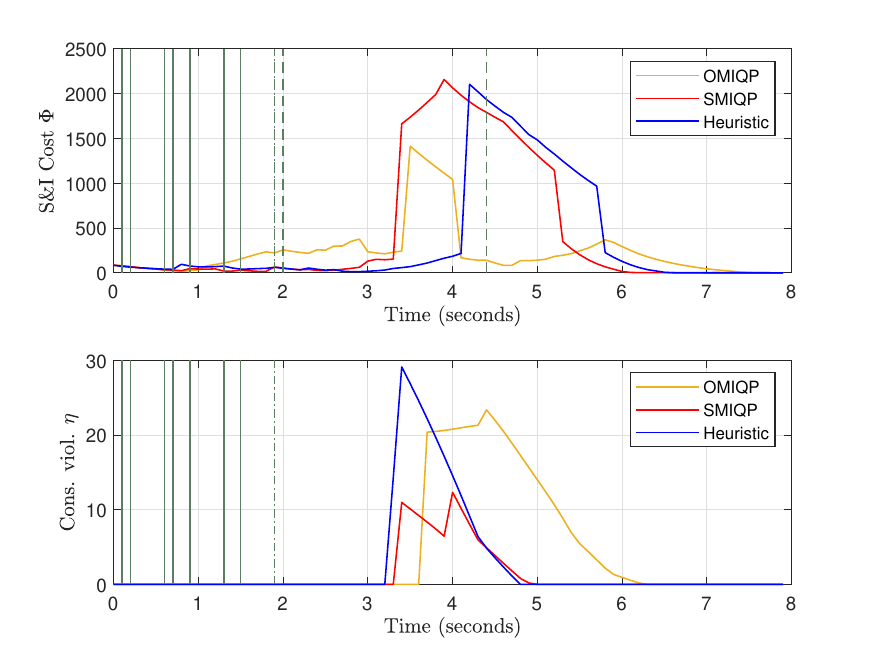}
        \end{minipage}
        \caption{Crossing order (left), closed-loop cost evolution, and constraint violation (right) from the \emph{nominal} test for all methods. The top left subfigure shows the crossing order of the benchmark simplified MIQP (SMIQP) (dashed lines), and the original MIQP (OMIQP) (solid lines), while the bottom left one is from the heuristic/heur. (solid lines). Similarly, the top right subfigure shows the costs, while the lower one displays the constraint violation. The vertical dashed, solid, and dashed-dotted green lines indicate the reordering timings $\tau$ of the OMIQP, SMIQP, and heuristic, respectively}
    \label{fig:e1-oj}			
    \end{center}
\end{figure*}     

\begin{figure*}[]
    \begin{center}
        \begin{minipage}{.5\textwidth}
            \centering
            \includegraphics[width=\linewidth]{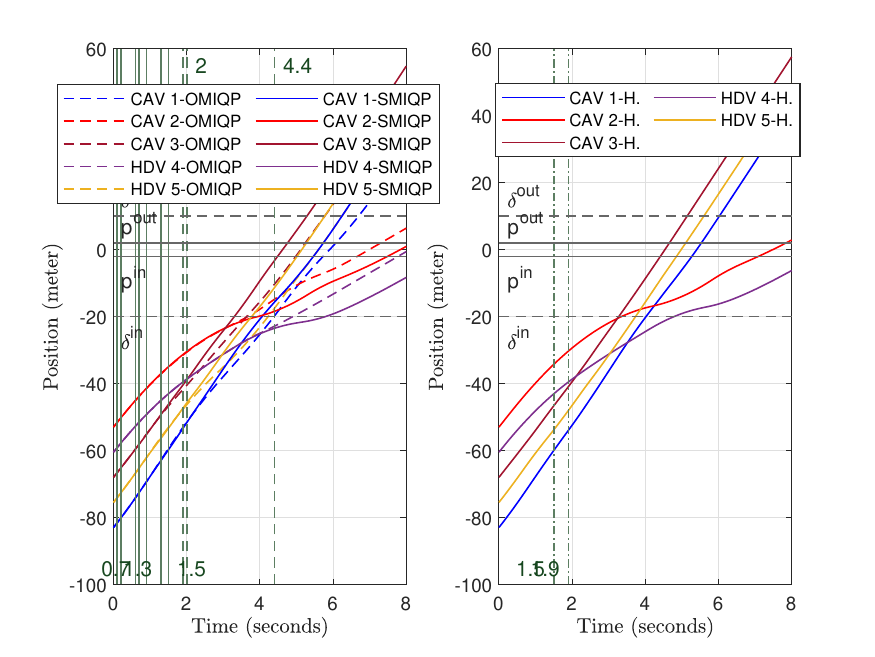}
        \end{minipage}%
        \begin{minipage}{.5\textwidth}
            \centering
            \includegraphics[width=\linewidth]{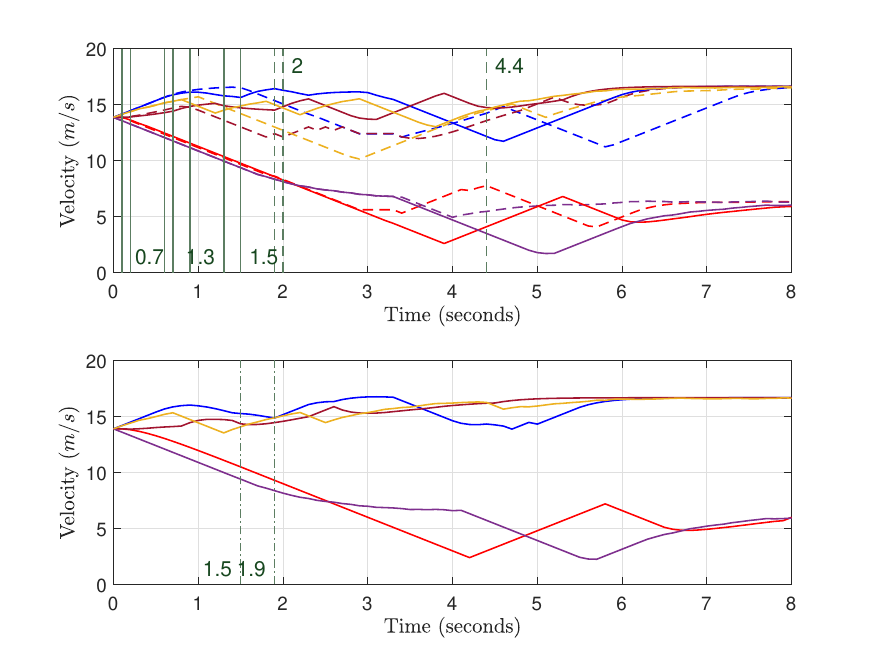}
        \end{minipage}
        \caption{Position/trajectory (left) and velocity (right) profiles from the \emph{nominal} test for all methods. The most left subfigure is relative to the benchmark simplified MIQP (SMIQP) (solid lines), and the original MIQP / OMIQP (dashed lines), while the next subfigure is relative to the heuristic. Similarly, the top right subfigure shows the velocity of the MIQPs, while the bottom right one is from the heuristic. The legends are applied to all subfigures. The vertical green dashed, solid, and dashed-dotted lines indicate reordering timings $\tau$ of the OMIQP, SMIQP, and heuristic, respectively.}
    \label{fig:e1-pv}			
    \end{center}
\end{figure*}   
%

\begin{table*}[]
\centering
\caption{Performance comparison of all methods for nominal scenario}
\begin{tabular}{|l|l|l|l|l|l|r|r|}
\hline
\multicolumn{1}{|c|}{\textbf{Methods}} & \multicolumn{1}{c|}{\textbf{\begin{tabular}[c]{@{}c@{}} $|\boldsymbol{\tau}|$\end{tabular}}} & \multicolumn{1}{c|}{\textbf{\begin{tabular}[c]{@{}c@{}} Total cost \\ $ \Phi^{\mathrm{cl}}$ \end{tabular}}} & \multicolumn{1}{c|}{\textbf{\begin{tabular}[c]{@{}c@{}} S\&I cost \\ $\Phi^{\mathrm{cl, S\&I}}$  \end{tabular}}} & \multicolumn{1}{c|}{\textbf{\begin{tabular}[c]{@{}c@{}} $\boldsymbol{\eta}^{\mathrm{max}}$ [m]\end{tabular}}} & \multicolumn{1}{c|}{\textbf{\begin{tabular}[c]{@{}c@{}} $\textbf{u}^{\mathrm{RMS}}$ {[}m/s$^2${]}\end{tabular}}} & \multicolumn{1}{c|}{\textbf{\begin{tabular}[c]{@{}c@{}} $\textbf{t}^{\mathrm{max}}$ {[}s{]}\end{tabular}}} & \multicolumn{1}{c|}{\textbf{\begin{tabular}[c]{@{}c@{}}Times faster \\ than OMIQP\end{tabular}}} \\ \hline
\textbf{OMIQP}         & 7                                                                                             & 133812 																			  & 18531                                                                                   & 7.21                                                                                       & 2.37                                                                                                      & 234.03                                                                                            & N/A                                                                                                   \\ \hline
\textbf{SMIQP}       & 6                                                                                             & 47279 																				  & 35545                                                                                   & 3.28                                                                                        & 2.16                                                                                                      & 156.50                                                                                             & 1.49                                                                                                  \\ \hline
\textbf{Heuristic}                 & 2                                                                                             & 73117 																				  & 27853                                                                                   & 12.62                                                                                        & 1.95                                                                                                      & 1.14                                                                                              & 203.50                                                                                                 \\ \hline
\end{tabular}
\label{tab:case-1-a}
\end{table*}

\begin{figure}[]
    \begin{center}
    \graphicspath{{./figures/}}
    \includegraphics[scale = 0.5]{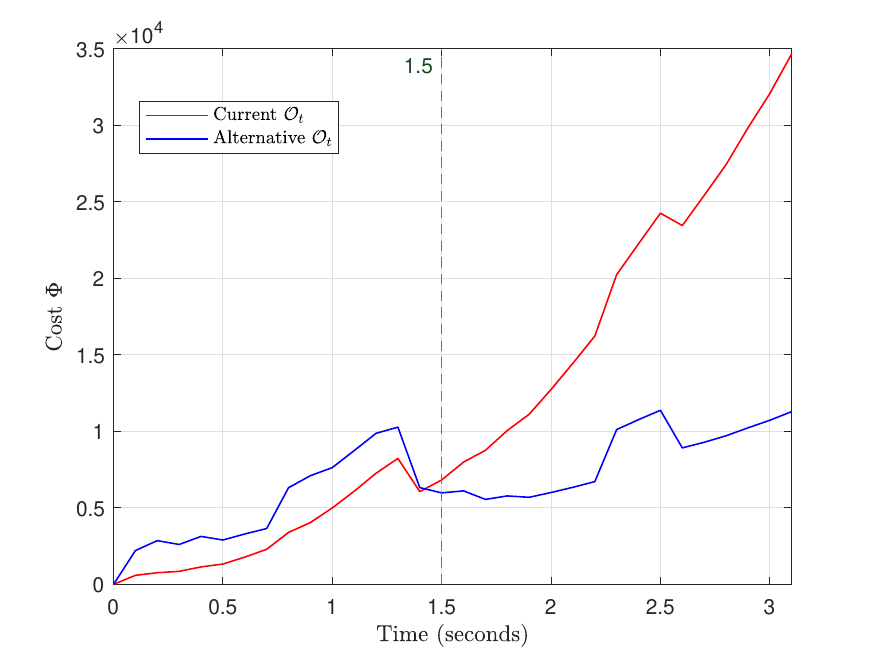}    
        \caption{Cost-to-go evolution of the current and alternative QPs solved inside the heuristic}
        \label{fig:j-comp}
    \end{center}
\end{figure}

We first compare the two MIQP formulations, i.e., OMIQP~\eqref{eq:miqp-relax}, SMIQP~\eqref{eq:miqp-simple}. To solve the MIQPs within a reasonable time, we limit their iterations to $1.7\times10^4$ and perform only a single simulation. To make a fair comparison between the MIQPs and the heuristic, the consistency check is not used here, i.e., $n^\text{max} = 0$.

Figure~\ref{fig:e1-oj} depicts the crossing order (left) and cost (right) yielded by the simulation. It can be observed that all vehicles start moving in the crossing order defined by their initial positions, i.e., $\mathcal{O}_{t=0} = [o_{1,0}, o_{2,0}, o_{3,0}]^{\top} = [2,3,1]^{\top} $. In the top left subfigure, it can be observed that both MIQPs' crossing orders chatter a few times between different combinations at the two seconds of the simulation, as one can see in the vertical green lines indicating $\tau^{\mathrm{OMIQP}}_{1-4}$ and $\tau^{\mathrm{SMIQP}}_{1-5}$ until the order finally settles. This issue is the consequence of the additive perturbation, which, in case two orders have very similar costs, makes the MIQP solvers alternate between one order and the other. 

HDV $4$ is set to slow down, as depicted in Figure~\ref{fig:e1-pv} on the right, and therefore platoons $2$ and $3$ also have to slow down, until the order is changed by swapping platoons $2$ and $3$. SMIQP eventually settles at $\mathcal{O}_{t} = [3,2,1]^{\top} $ at time $\tau_6^{\mathrm{SMIQP}} = 1.3$ s (top left subfigure of Figure~\ref{fig:e1-oj}). The heuristic follows the same reordering at $\tau_1^{\mathrm{H}} = 1.5$ s (bottom left subfigure) without any chattering. The difference in the switching time between the methods is caused by the approximate nature of the timing binaries $\boldsymbol{\rho}$ used by the heuristic. When solving FO-QPs~\eqref{eq:miqp-fo} of the current and alternative orders in the heuristic, $\hat{\boldsymbol{\rho}}$ are parameters approximated from the predicted trajectories at previous $t_k$, while in SMIQP, they are optimization decision variables which are solved together with the vehicle trajectories. Therefore, their values are not always the same, and this may create a difference in the reordering timing. OMIQP eventually also yields the same order at a later time $\tau^{\mathrm{OMIQP}}_4 = 2.0$ s. This is due to the higher complexity of the OMIQP formulation, which entails that the optimal solution cannot be obtained within the imposed iteration limit.

The same pattern can be observed for the three algorithms, as well as for the next switching time, which occurs at time $\tau^{\mathrm{SMIQP}}_7 = 1.5$ s for SMIQP and at time $\tau^{\mathrm{H}}_2 = 1.9$ s for the heuristic. They converge to the final order $\mathcal{O}_{t} = [3,1,2]^{\top}$. The final order is also obtained by OMIQP at a later time $\tau^{\mathrm{OMIQP}}_5 = 4.4$ s, which verifies the SMIQP and heuristics final order decisions. The motivation for reordering decisions is illustrated in Figure~\ref{fig:j-comp} which presents the cost-to-go comparison run inside the heuristic between the QPs of current (first) and alternative orders. If the current order were kept, the cost would continue to increase, so that a reordering, e.g. at $\tau^{\mathrm{H}}_1$, yields a lower cost. 

In Table~\ref{tab:case-1-a}, one can see that the sum of the closed-loop state reference tracking and input (S\&I) cost~$\Phi^{\mathrm{cl,S\&I}}$ of our heuristic is higher than the one of OMIQP, but lower than the one of SMIQP. SMIQP yields the highest S\&I cost here because it needs to minimize future constraint violations. Consequently, it triggers reordering early which requires more use of inputs and reference speed deviation as a trade-off for the platoons to swap their positions. This is encountered, for example, by CAV $2$ that significantly decelerates as shown in Figure~\ref{fig:e1-pv}. In Figure~\ref{fig:e1-oj}, we see that this results in the S\&I cost of SMIQP being the highest between $t=3.4 - 4$ s. On the other hand, it yields the lowest maximum constraint violation~$\boldsymbol{\eta}^{\mathrm{max}}$, which, as depicted in the left subfigure of Figure~\ref{fig:e1-pv}, occurs around the entrance safety margin $\delta^{\mathrm{in}}$ and before CZ hence practically safe. The early reorderings also imply that the SMIQP's sum of total cost $\Phi^{\mathrm{cl}}$, which includes the slack terms, is the lowest among the others. 

In Figure~\ref{fig:e1-oj}, it is seen that the heuristic has relatively the same S\&I cost evolution as SMIQP, but it is slightly delayed. This yields lower cost which is due to the lower use of inputs and reference deviation for swapping at later times, but consequently, the maximum constraint violation and total cost are higher than those of SMIQP. 

Further, we can see that OMIQP has the lowest S\&I cost. This is due to the fact that the platoons are trying to swap even though reordering decisions come late as the solver does not reach convergence. Hence, it can be seen in Figure~\ref{fig:e1-oj}, that in around $t = 1-3$ s, OMIQP S\&I cost is the highest but the required inputs and reference deviation for swapping later are eventually lowest. However, one can see, as in the bottom subfigure, that this decision results in more use of slacks that significantly contribute to the total cost resulting in the OMIQP having the highest $\Phi^{\mathrm{cl}}$. 

In the average (RMS) control input $\textbf{u}^{\mathrm{RMS}}$ exhibited by each CAV, it can be seen that the gap between the SMIQP and heuristic here is much smaller than the gap in the cost, which shows that the heuristic yields approximately close control input to SMIQP. Due to the imposed iteration bound, as expected, OMIQP yields a slightly larger RMS input. 

Figure~\ref{fig:e1-pv} displays the positions and velocities obtained by MIQPs and our heuristic. One can observe from the velocity curves on the right figure that the platoon $3$ of the SMIQP and the heuristic have to marginally slow down due to the current order until reordering is executed. It is noteworthy to observe that SMIQP and the heuristic yield very similar position and velocity profiles. Instead, due to later reordering times, OMIQP forces CAV $3$ and $1$ to slow down for a longer time and, consequently, aside from platoon $2$, all platoons enter the CZ at a later time, compared to SMIQP and our heuristic. 
 
The main advantage of the heuristic over the MIQPs is shown in the (worst-case) computation time reported in Table~\ref{tab:case-1-a}: the heuristic is about $203$ times faster to solve than OMIQP and also about a hundred times faster than SMIQP. SMIQP is about 1.5 times faster than OMIQP, but that does not make any significant difference in the context of time-tractable coordination.

\subsubsection{Simulation with random HDV input bounds}
\label{subsubsec:exp-1-b}
%
\begin{table*}[]
\centering
\caption{Average performance of the heuristic for nominal scenario with different input bounds}
\begin{tabular}{|l|l|l|l|l|l|l|r|}
\hline
\multicolumn{1}{|c|}{\textbf{Methods}} & \multicolumn{1}{c|}{\textbf{\begin{tabular}[c]{@{}c@{}} $|\boldsymbol{\tau}|$\end{tabular}}} & \multicolumn{1}{c|}{\textbf{\begin{tabular}[c]{@{}c@{}}Average \\ $\tau$\end{tabular}}} & \multicolumn{1}{c|}{\textbf{\begin{tabular}[c]{@{}c@{}}Total cost \\ $\Phi^{\mathrm{cl}}$ \end{tabular}}} & \multicolumn{1}{c|}{\textbf{\begin{tabular}[c]{@{}c@{}} S\&I cost \\ $\Phi^{\mathrm{cl, S\&I}}$  \end{tabular}}} & \multicolumn{1}{c|}{\textbf{\begin{tabular}[c]{@{}c@{}} $\boldsymbol{\eta}^{\mathrm{max}}$ [m]\end{tabular}}} & \multicolumn{1}{c|}{\textbf{\begin{tabular}[c]{@{}c@{}} $\textbf{u}^{\mathrm{RMS}}$ {[}m/s$^2${]}\end{tabular}}} & \multicolumn{1}{c|}{\textbf{\begin{tabular}[c]{@{}c@{}} $\textbf{t}^{\mathrm{max}}$ {[}s{]}\end{tabular}}} \\ \hline
\textbf{Heuristic}               & 2                                                                                             & 1.96                                                                                  & 92683  & 30187                                                             &   15.66            & 1.97                                                                                    & 1.28                                                                                              \\ \hline
\end{tabular}
\label{tab:case-1-b}
\end{table*} 
In this second type of simulation, the random HDV input bounds are sampled from a uniform distribution within $10\%$ of the nominal ones $(\pm 3~\text{m/s}^{2})$. The consistency check is used with $n^\text{max}=3$, as shown in Table~\ref{tab:param}. As mentioned before, MIQPs are not executed here due to their long computation times. 

We provide in Table~\ref{tab:case-1-b} the values averaged over ten simulations. One can see that even though the input bounds are different in each simulation, the heuristic performs reordering only twice. The consistency check plays an important role here, as it makes sure that an order swap only occurs when it is advantageous enough, therefore mitigating the chattering behavior of the MIQPs as observed in the previous part~\ref{subsubsec:exp-1-a}. Moreover, the maximum constraint violation, total, and S\&I costs here are slightly higher than the heuristic in the previous simulation~\ref{subsubsec:exp-1-a} but the average acceleration remains almost the same. In terms of computation time, one can see that the heuristic is consistently faster compared to the MIQPs in the previous part.

%
\subsection{Disturbance Scenario}
\label{subsec:exp-2}

\begin{figure}[]
    \begin{center}
    \graphicspath{{./figures/}}
    \includegraphics[scale = 0.5]{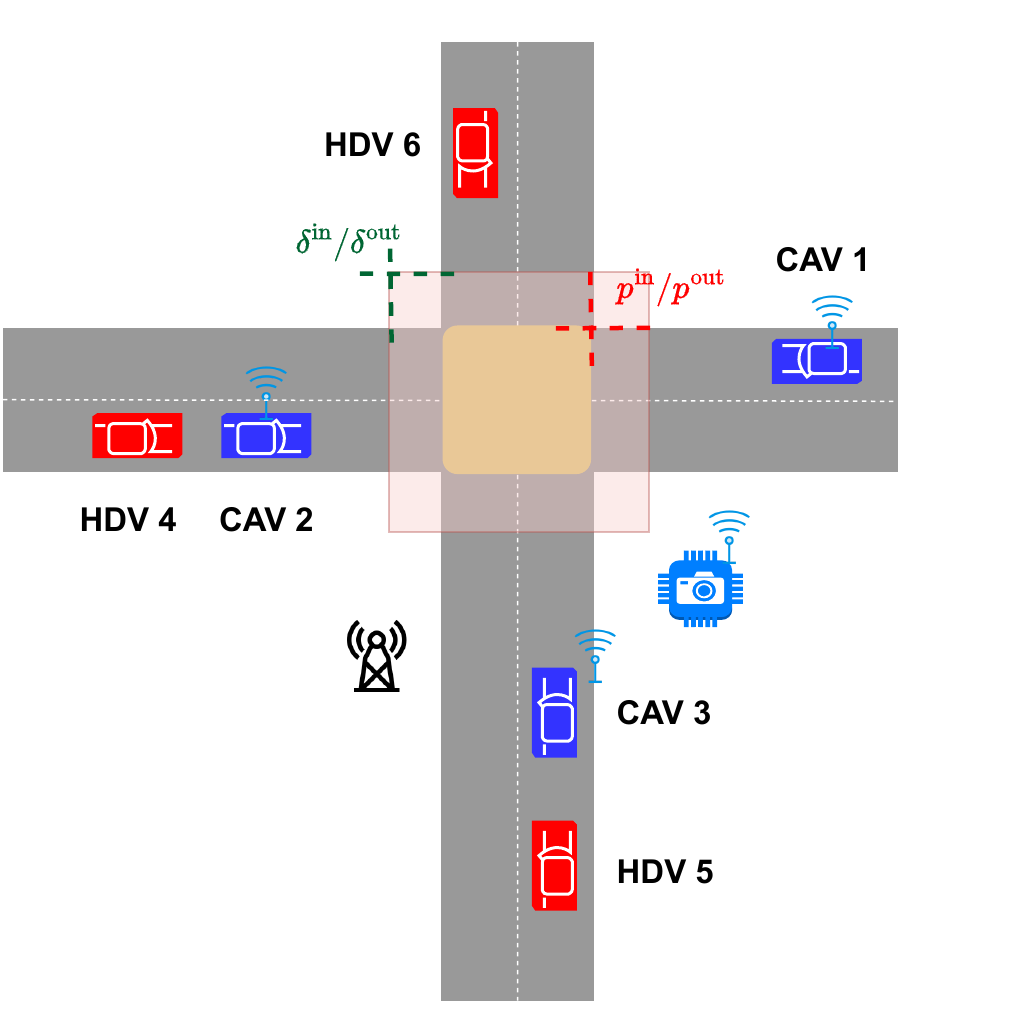}    
        \caption{Vehicle configuration for \emph{disturbance} case}
        \label{fig:case-2}
    \end{center}
\end{figure}
This scenario includes an additional \emph{leading} HDV $6$ alongside existing vehicles, as shown in Figure~\ref{fig:case-2}, acting as a \emph{disturbance} to the platoons' coordination. Without any CAV in front that can regulate HDV $6$, all CAVs are imposed an additional safety constraint by the intersection manager (IM): HDV $6$ must occupy the intersection first. HDV 6 tracks a constant speed $v^\text{ref}_i = 60$ km/h and is initially positioned at $86,25$ meters from the center of the CZ with an initial velocity of $50$ km/h. It is also modeled using the switching model~\eqref{eq:hdv-cd}. The initial position of CAV $2$ is shifted to $60$ m from the center of the CZ, and the rest of the vehicles are configured with the same sequence from CAV $2$ as in the previous nominal case, also with $7.5$ m gap each. Here, HDV $4$ does not slow down and instead follows $v^\text{nom}$. The prediction and simulation horizon are selected as $T^{\mathrm{p}} = 3.5$ s and $T^{\mathrm{sim}} = 10$ s. Finally, we select $\bar{d}=8$ m and $d^{\mathrm{ref}}=10$ m.

In these simulations, we drop OMIQP as the scenario is more complex than before, and the solver does not manage to converge in a reasonable time. For SMIQP, the limit of iterations imposed is $10^4$ iterations. 

\subsubsection{Simulation against MIQPs}
\label{subsubsec:exp-2-a}

\begin{table*}[]
\centering
\caption{Performance comparison of SMIQP and heuristic for disturbance scenario}
\begin{tabular}{|l|l|l|l|l|l|r|r|}
\hline
\multicolumn{1}{|c|}{\textbf{Methods}} & \multicolumn{1}{c|}{\textbf{\begin{tabular}[c]{@{}c@{}} $|\boldsymbol{\tau}|$\end{tabular}}} & \multicolumn{1}{c|}{\textbf{\begin{tabular}[c]{@{}c@{}} Total cost \\ $\Phi^{\mathrm{cl}}$ \end{tabular}}} & \multicolumn{1}{c|}{\textbf{\begin{tabular}[c]{@{}c@{}} S\&I cost \\$\Phi^{\mathrm{cl, S\&I}}$  \end{tabular}}} & \multicolumn{1}{c|}{\textbf{\begin{tabular}[c]{@{}c@{}} $\boldsymbol{\eta}^{\mathrm{max}}$ [m]\end{tabular}}} & \multicolumn{1}{c|}{\textbf{\begin{tabular}[c]{@{}c@{}} $\textbf{u}^{\mathrm{RMS}}$ {[}m/s$^2${]}\end{tabular}}} & \multicolumn{1}{c|}{\textbf{\begin{tabular}[c]{@{}c@{}} $\textbf{t}^{\mathrm{max}}$ {[}s{]}\end{tabular}}} & \multicolumn{1}{c|}{\textbf{\begin{tabular}[c]{@{}c@{}}Times faster \\ than SMIQP\end{tabular}}} \\ \hline
\textbf{SMIQP}       & 14                                                                                            & 147840 & 106109                                                                                   & 11.79                                                                                        & 2.42                                                                                                      & 752.64                                                                                             & N/A                                                                                                  \\ \hline
\textbf{Heuristic}                 & 2                                                                                             & 157620 & 84268                                                                                      & 6.43                                                                                        & 2.37                                                                                                      & 1.80                                                                                              & 417.97                                                                                                 \\ \hline
\end{tabular}
\label{tab:case-2-a}
\end{table*}
%

\begin{figure*}[]
    \begin{center}
        \begin{minipage}{.5\textwidth}
            \centering
            \includegraphics[width=\linewidth]{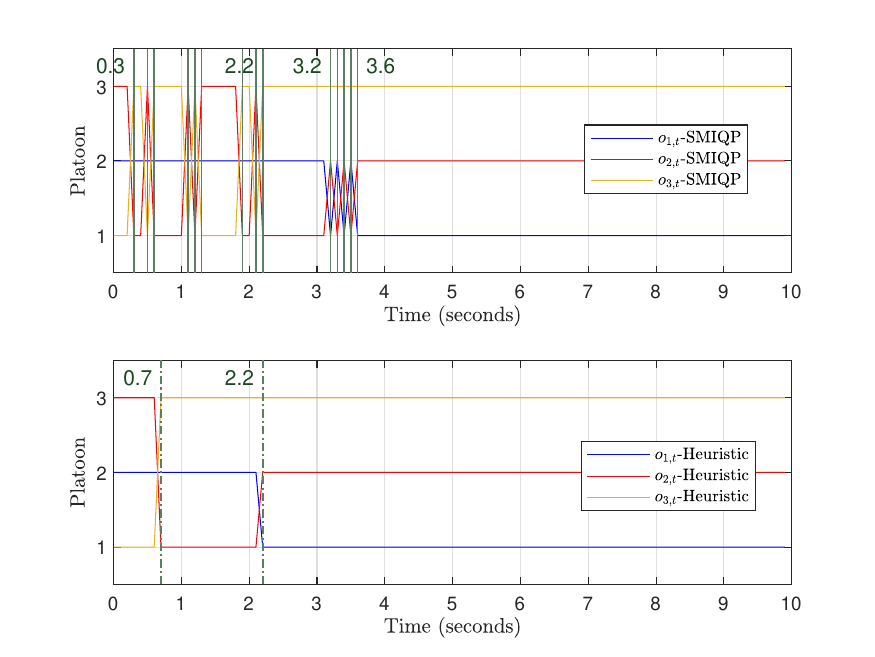}
        \end{minipage}%
        \begin{minipage}{.5\textwidth}
            \centering
            \includegraphics[width=\linewidth]{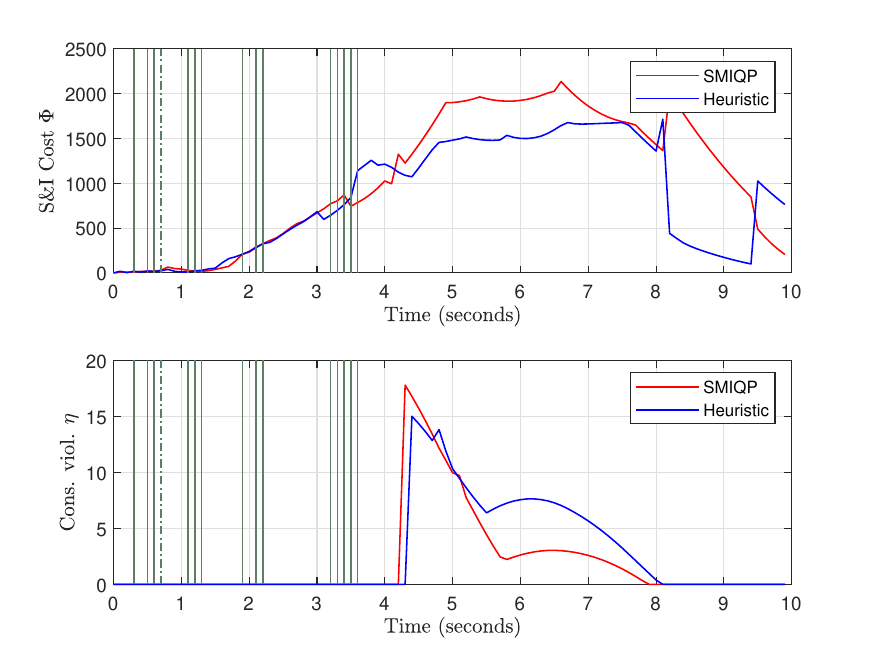}
        \end{minipage}
        \caption{Crossing order (left) and closed-loop cost evolution (right) from the \emph{disturbance} test for all methods. The top left subfigure shows the crossing order of the benchmark simplified MIQP (SMIQP), while the bottom left one is from the heuristic (H). The top right subfigure shows the costs while the bottom right one displays the constraint violations. The solid and dashed-dotted vertical green lines indicate reordering timings $\tau$ that are relative to SMIQP and heuristic, respectively.}
    \label{fig:e2-oj}			
    \end{center}
\end{figure*}     

\begin{figure*}[]
    \begin{center}
        \begin{minipage}{.5\textwidth}
            \centering
            \includegraphics[width=\linewidth]{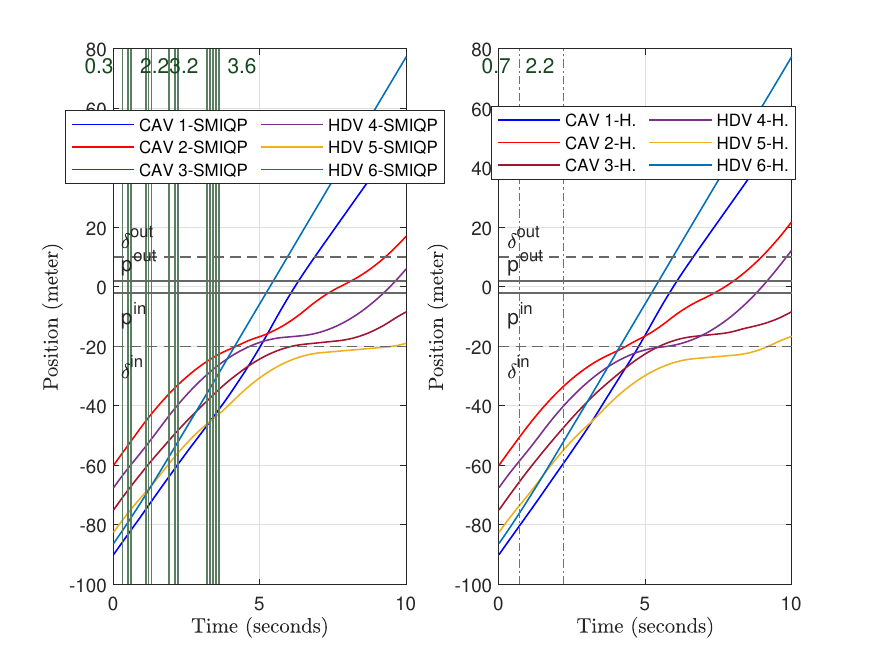}
        \end{minipage}%
        \begin{minipage}{.5\textwidth}
            \centering
            \includegraphics[width=\linewidth]{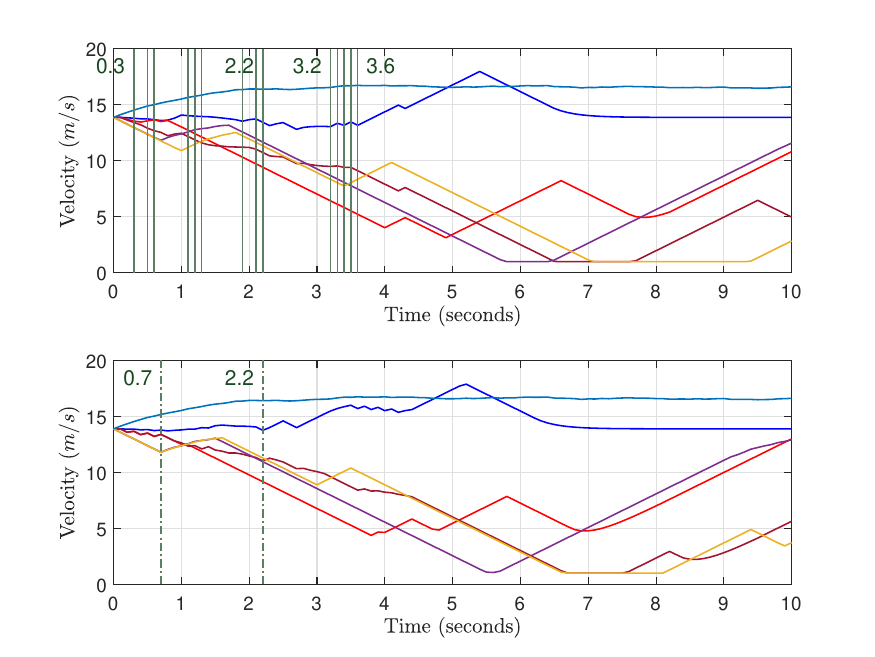}
        \end{minipage}
        \caption{Position/trajectory (left) and velocity (right) profiles from the \emph{disturbance} test for all methods. The left subfigure is relative to the benchmark simplified MIQP / SMIQP, while the central one is relative to the heuristic (H). Similarly, the top right subfigure shows the velocity of the benchmark, while the bottom right one is relative to the heuristics. The color and lines from the legends are applied to all subfigures. The solid and dashed-dotted vertical green lines indicate reordering timings $\tau$ that are relative to SMIQP and heuristic, respectively.}
    \label{fig:e2-pv}			
    \end{center}
\end{figure*}  

As can be seen in Figures~\ref{fig:e2-oj} and~\ref{fig:e2-pv}, the initial order is dictated by the initial positions of the vehicles, that is, $\mathcal{O}_{0} = [2,3,1]^{\top} $. Due to the additional safety constraint against HDV 6, this causes the platoons to gradually decelerate, except for platoon (CAV) $1$, as seen in the velocity profiles of the figure. In the beginning, platoon $2$ is seen to have the highest deceleration among the other platoons as it is the one currently closest to the intersection and needs to slow down to let HDV 6 come first. 

As platoons $2$ and $3$ slow down, this forces the tail HDVs behind them to also decelerate. This action is necessary to prevent the possibility of HDVs $4$ and $5$ from occupying the CZ at the same time as HDV $6$, which may result in a potential deadlock or collision. Platoon $1$, which is a one-vehicle platoon, is then given a chance to overtake both of them gradually, as can be seen in Figure~\ref{fig:e2-pv}. 

In Figure~\ref{fig:e2-oj} in the upper left, it can be observed that SMIQP chatters between $\mathcal{O}_{t} = [2,3,1]^{\top} $ and $\mathcal{O}_{t} = [2,1,3]^{\top} $ several times starting at $\tau^{\mathrm{SMIQP}}_{1} = 0.3$ s. The order is eventually settled to the latter at $\tau^{\mathrm{SMIQP}}_{9} = 2.2$ s. On the other hand, the heuristic requires only a single swap at $\tau^{\mathrm{H.}}_{1} = 0.7$ s to do so. Finally, the crossing order is further modified to $\mathcal{O}_{t} = [1,2,3]^{\top} $ by both approaches. In this case, SMIQP switches last, with some chattering as well, which is due to the imposed iteration limit as observed in the solver outputs.

As in the previous case~\ref{subsec:exp-1}, the heuristic here can avoid the chattering issue without the consistency check. The two approaches converge to a similar final order $\mathcal{O}_{t} = [1,2,3]^{\top} $. The approximately similar RMS input indicates that their trajectories are reasonably close, as observed in Figure~\ref{fig:e2-pv}, and the cost difference is dominated by slack values.

Furthermore, the S\&I cost $\Phi^{\mathrm{cl,S\&I}}$ of the heuristic is lower than that of SMIQP, but the total cost $\Phi^{\mathrm{cl}}$ (with the slack terms) is the opposite. This indicates the same trade-off seen in the previous scenario, where the early reordering performed by SMIQP yields lower constraint violation at the price of a slightly higher S\&I cost and RMS acceleration. This can be concluded from Figure~\ref{fig:e2-oj} where cost evolution from the methods here are generally similar, but it starts to differ around $t=4.5$ s where due to the early reordering the cost is slightly higher for SMIQP. Due to the taking of subsequent reorderings, the heuristic has a slightly higher violation from $t = 5$ s. Additionally, one can see that the worst-case $\boldsymbol{\eta}^{\mathrm{max}}$ of SMIQP is slightly higher than that of the heuristic which in this scenario is due to the iteration limit. 

As seen in the previous scenario~\ref{subsubsec:exp-1-a}, the approximation of timing binaries in the heuristic leads to differences in reordering. However, the heuristic is shown to yield significantly smaller computational times, about $400$ times faster, at the price of a marginal constraint violation increase, but with a better crossing order consistency and close to optimal trajectories.
\subsubsection{Simulation with random HDV inputs}

\begin{table*}[]
\centering
\caption{Average performance comparison for disturbance scenario - $2^{\mathrm{nd}}$ test}
\begin{tabular}{|l|l|l|l|l|l|l|r|}
\hline
\multicolumn{1}{|c|}{\textbf{Methods}} & \multicolumn{1}{c|}{\textbf{\begin{tabular}[c]{@{}c@{}} $|\boldsymbol{\tau}|$\end{tabular}}} & \multicolumn{1}{c|}{\textbf{\begin{tabular}[c]{@{}c@{}}Average \\ $\tau$\end{tabular}}} & \multicolumn{1}{c|}{\textbf{\begin{tabular}[c]{@{}c@{}} Total cost \\ $\Phi^{\mathrm{cl}}$ \end{tabular}}} & \multicolumn{1}{c|}{\textbf{\begin{tabular}[c]{@{}c@{}} S\&I cost \\$\Phi^{\mathrm{cl, S\&I}}$  \end{tabular}}} & \multicolumn{1}{c|}{\textbf{\begin{tabular}[c]{@{}c@{}} $\boldsymbol{\eta}^{\mathrm{max}}$ [m]\end{tabular}}} & \multicolumn{1}{c|}{\textbf{\begin{tabular}[c]{@{}c@{}} $\textbf{u}^{\mathrm{RMS}}$ {[}m/s$^2${]}\end{tabular}}} & \multicolumn{1}{c|}{\textbf{\begin{tabular}[c]{@{}c@{}} $\textbf{t}^{\mathrm{max}}$ {[}s{]}\end{tabular}}}  \\ \hline
\textbf{Heuristic}               & 2, 4                                                                                             & 2.21                                                                                    & 166900 & 90980                                                              & 10.30               & 2.39                                                                                    & 1.65                                                                                              \\ \hline
\end{tabular}
\label{tab:case-2-b}
\end{table*} 
In this comparison, we further conduct ten different simulations with randomized HDV input bounds for the heuristic, with a similar vehicle configuration, e.g., initial conditions, etc., as in the previous simulation~\ref{subsubsec:exp-2-a}. The input bounds are generated analogously to those of~\ref{subsubsec:exp-1-b}. The consistency check is applied here with $n^{\mathrm{max}} = 3$. Also, SMIQP is not executed here due to its long computation time. 

The resulting average values of the simulations are presented in Table~\ref{tab:case-2-b}. One can see that the heuristic yields $2$ and $4$ reordering occasions. Among the ten simulations, only one instance has $4$ reordering here, and the others are $2$. This shows that while chattering is not eliminated completely as a result of having perturbation, useless order changes are kept within a reasonable number and it only occurs to one sample out of ten.

From the costs and maximum constraint violation perspectives, one can see that the values are slightly higher than the previous simulation in~\ref{subsubsec:exp-2-a}. Also, the worst-case computation time and RMS of CAV acceleration, shown in Table~\ref{tab:case-2-b}, remain small. These remarks indicate consistency of the heuristic's performance throughout simulations in a scenario against a leading HDV.

%
\section{CONCLUSIONS}
\label{sec:conclusions}

We proposed a heuristic algorithm designed to coordinate both CAVs and HDVs at unsignalized intersections. Our algorithm is able to efficiently handle dynamic reordering problems that may arise due to changes in HDVs' trajectories. Obtaining the optimal crossing order and acceleration profiles requires solving computationally expensive MIQPs, which are not real-time feasible. Our heuristic overcomes this challenge by combining a problem-tailored future constraint violation/consistency check and a cost comparison strategy. The consistency check allows us to restrict reordering actions to a smaller subset of platoons, therefore mitigating subproblems branching complexity. The cost comparison eventually decides whether an order change takes place or not. Simulation results demonstrate that our heuristic is orders of magnitude faster than solving MIQPs and has better order consistency, at the price of a marginal performance degradation.

Future work will further develop numerical aspects of the heuristic to improve its performance while addressing more challenging scenarios, e.g., involving a continuous flow of vehicles simulated using traffic simulators and considering nonlinear dynamical models.


%
\section*{REFERENCES}

\printbibliography[heading=none]


 




\vfill

\end{document}